\documentclass[twocolumn,showpacs,preprintnumbers,amsmath,amssymb,latexsym,prl,footinbib,floatfix,superscriptaddress]{revtex4-2}

\usepackage{amsmath,amssymb,latexsym,slashed,multirow}

\usepackage{afterpage}
\usepackage{graphicx}
\usepackage{xcolor}
\newcommand{\f}[2]{\frac{#1}{#2}}

\newcommand{\Dmeas}{\mathcal{D}}

\begin{document}

\title{Equation of state of a hot-and-dense quark gluon plasma: \\
lattice simulations at real $\mu_B$ vs. extrapolations
}

\author{Szabolcs Bors\'anyi} \affiliation{Department of Physics,
  Wuppertal University, Gaussstr.\ 20, D-42119, Wuppertal, Germany}

\author{Zolt\'an Fodor} 
  \affiliation{Department of Physics, Wuppertal University, Gaussstr.\ 20, D-42119, Wuppertal, Germany}
  \affiliation{Pennsylvania State University, Department of Physics, State College, Pennsylvania 16801, USA}
  \affiliation{J{\"u}lih Supercomputing Centre, Forschungszentrum J{\"u}lich, D-52425 J{\"u}lich, Germany}
  \affiliation{ELTE E\"otv\"os Lor\'and University, Institute for
  Theoretical Physics, P\'azm\'any P\'eter s\'et\'any 1/A, H-1117, Budapest,
  Hungary}
  \affiliation{Physics Department, USDA, San Diego, CA 92093, USA}

\author{Matteo Giordano}
\affiliation{ELTE E\"otv\"os Lor\'and University, Institute for
  Theoretical Physics, P\'azm\'any P\'eter s\'et\'any 1/A, H-1117, Budapest,
  Hungary}

\author{Jana N. Guenther}
\affiliation{Department of Physics, Wuppertal
  University, Gaussstr.\ 20, D-42119, Wuppertal, Germany}

\author{Sandor D.\ Katz}
\affiliation{ELTE E\"otv\"os Lor\'and University, Institute for
  Theoretical Physics, P\'azm\'any P\'eter s\'et\'any 1/A, H-1117, Budapest,
  Hungary}
\affiliation{
  MAT-ELTE Theoretical Physics Research Group,
  P\'azm\'any P\'eter s\'et\'any 1/A, H-1117 Budapest, Hungary.}

\author{Attila P\'asztor}
\email{Corresponding author: apasztor@bodri.elte.hu}
\affiliation{ELTE E\"otv\"os Lor\'and University, Institute for
  Theoretical Physics, P\'azm\'any P\'eter s\'et\'any 1/A, H-1117, Budapest,
  Hungary}

\author{Chik Him Wong} \affiliation{Department of Physics, Wuppertal
  University, Gaussstr.\ 20, D-42119, Wuppertal, Germany}

\begin{abstract}
The equation of state of the quark gluon plasma is a key 
ingredient of heavy ion 
phenomenology. In 
addition to the traditional Taylor method, 
several novel approximation schemes have been 
proposed with the aim of calculating it at finite baryon density. 
In order to gain a pragmatic understanding of the limits of these schemes, we 
compare them to direct results at $\mu_B>0$, using 
reweighting techniques free from an overlap problem. 
We use 2stout improved staggered fermions with 8 time-slices 
and cover the entire RHIC BES range in the baryochemical potential, 
up to $\mu_B/T=3$.
\end{abstract}

\maketitle

\paragraph{Introduction}

The equation of state of strongly interacting matter under extreme conditions - high 
temperatures or baryon densities - plays a key role in many physical systems, 
such as the early Universe, heavy ion collisions and neutron stars. 
The most well established first-principles method to study 
the strongly coupled regime is lattice QCD~\cite{Montvay:1994cy}, that maps 
the path integral formulation of QCD to a classical statistical-mechanical 
system, which
can be simulated with Monte Carlo methods.
Indeed, many properties of strongly interacting 
matter at zero baryon density have been elucidated using this method, such as the 
crossover nature of the transition~\cite{Aoki:2006we}, the value of the transition 
temperature~\cite{Borsanyi:2010bp,Bazavov:2011nk} 
and the equation of state~\cite{Borsanyi:2010cj,Bazavov:2014pvz}.
Studies  
at finite baryon 
density are, however, hampered by the sign problem: the 
Boltzmann weights in the usual path integral 
representation become complex, 
making them unsuitable for importance sampling. Thus, most lattice 
results on the properties of hot-and-dense QCD matter rely on extrapolations from zero~\cite{Gavai:2003mf,Allton:2005gk,
  MILC:2008reg,Borsanyi:2011sw,Borsanyi:2012cr,
  Bellwied:2015lba,Ding:2015fca,Bazavov:2017dus,HotQCD:2018pds,
  Giordano:2019slo,Bazavov:2020bjn} 
or purely imaginary chemical potential~\cite{deForcrand:2002hgr,DElia:2002tig,DElia:2009pdy,
  Cea:2014xva,Bonati:2014kpa,Cea:2015cya,Bonati:2015bha,
  Bellwied:2015rza,DElia:2016jqh,Gunther:2016vcp,Alba:2017mqu,
  Vovchenko:2017xad,Bonati:2018nut,Borsanyi:2018grb,Bellwied:2019pxh,
  Borsanyi:2020fev}, two situations with no sign problem.

In spite of the difficulties, there has recently been considerable 
progress in the calculation of the equation of state 
of a hot-and-dense quark gluon plasma (QGP). First, Taylor 
coefficients of the pressure $p$ in the baryochemical potential
$\mu_B$ have been calculated up to fourth order in the 
continuum limit~\cite{Borsanyi:2012cr, Bellwied:2015lba, Bazavov:2017dus} and 
up to eighth order at finite lattice 
spacing~\cite{DElia:2016jqh,Borsanyi:2018grb,Bazavov:2020bjn} - albeit 
with rather large uncertainties for the sixth and eighth order coefficients. 
Second, several resummation schemes have been proposed  
for the Taylor 
expansion~\cite{Borsanyi:2021sxv,Mondal:2021jxk,Mitra:2022vtf,Bollweg:2022rps,
Borsanyi:2022qlh}, with a promise of better convergence 
properties. However, at the moment there is no
theoretical understanding of the convergence properties of these 
schemes.  
It is therefore important for phenomenological applications to 
determine the region of validity of these techniques.
This is the purpose of this work. 
To this end, we use some 
novel developments in simulation techniques 
for finite density QCD, concerning reweighting 
schemes where the pressure difference between the simulated and 
target theories  can be calculated without 
encountering heavy tailed 
distributions - i.e., an overlap problem. Recently, 
studies of this type have become feasible 
for improved lattice actions with physical quark masses 
for at least two schemes: sign 
reweighting, and phase reweighting 
~\cite{Giordano:2020roi, Borsanyi:2021hbk}.
These have the advantage of giving 
direct, reliable results at $\mu_B>0$ - provided that the 
exponentially difficult sign problem in the 
reweighting is dealt with by sufficient statistics. 
By comparing the equation of state calculated using phase 
reweighting with that coming from the Taylor expansion and 
its resummations, we can 
quantify the systematic bias of the different 
truncations of the different schemes,
giving unprecedented insight into the equation of state of 
the QGP. 

We simulate 2stout improved staggered 
fermions with physical quark masses on lattices with 8 
time slices - a discretization that is often used 
as the first or even the second point of continuum 
extrapolations of several thermodynamic quantities
~\cite{Aoki:2006we,Aoki:2006br,Borsanyi:2010cj,Bali:2011qj,
  Bali:2012zg,Borsanyi:2015yka,Bonati:2015bha,Brandt:2017oyy,
  Bonati:2018nut,DElia:2019iis}. The cut-off effects on the equation of state
  are moderate (after applying a tree-level 
  improvement~\cite{Borsanyi:2010cj}).
We reimplement all discussed extrapolation schemes using this same
setup. Thus, the differences in the predictions can only come from
the systematics of the extrapolation. 
Since we are mostly interested in a comparison of various methods for a statistical 
physics system, the physical volume can be chosen freely. We use a fixed aspect ratio
of $LT=2$, with $T$ the temperature and $L$ the spatial size of the box.
As we perform no finite volume scaling study, we must restrict ourselves to the 
study of the QGP equation of state, leaving that of the 
fate of the crossover transition to a later date.

\begin{figure*}[t]
  \centering
  \includegraphics[width=0.42\textwidth]{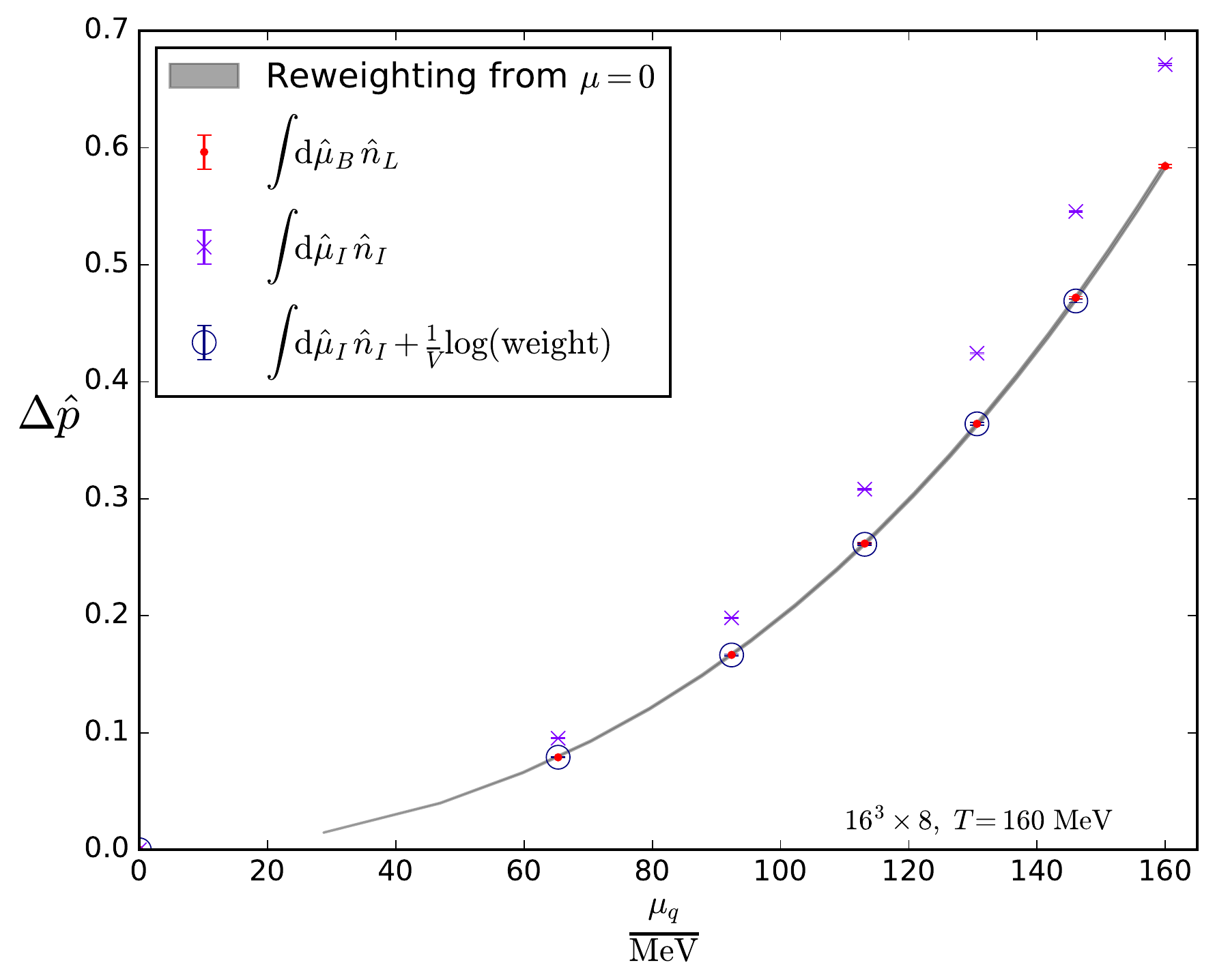}
  \includegraphics[width=0.42\textwidth]{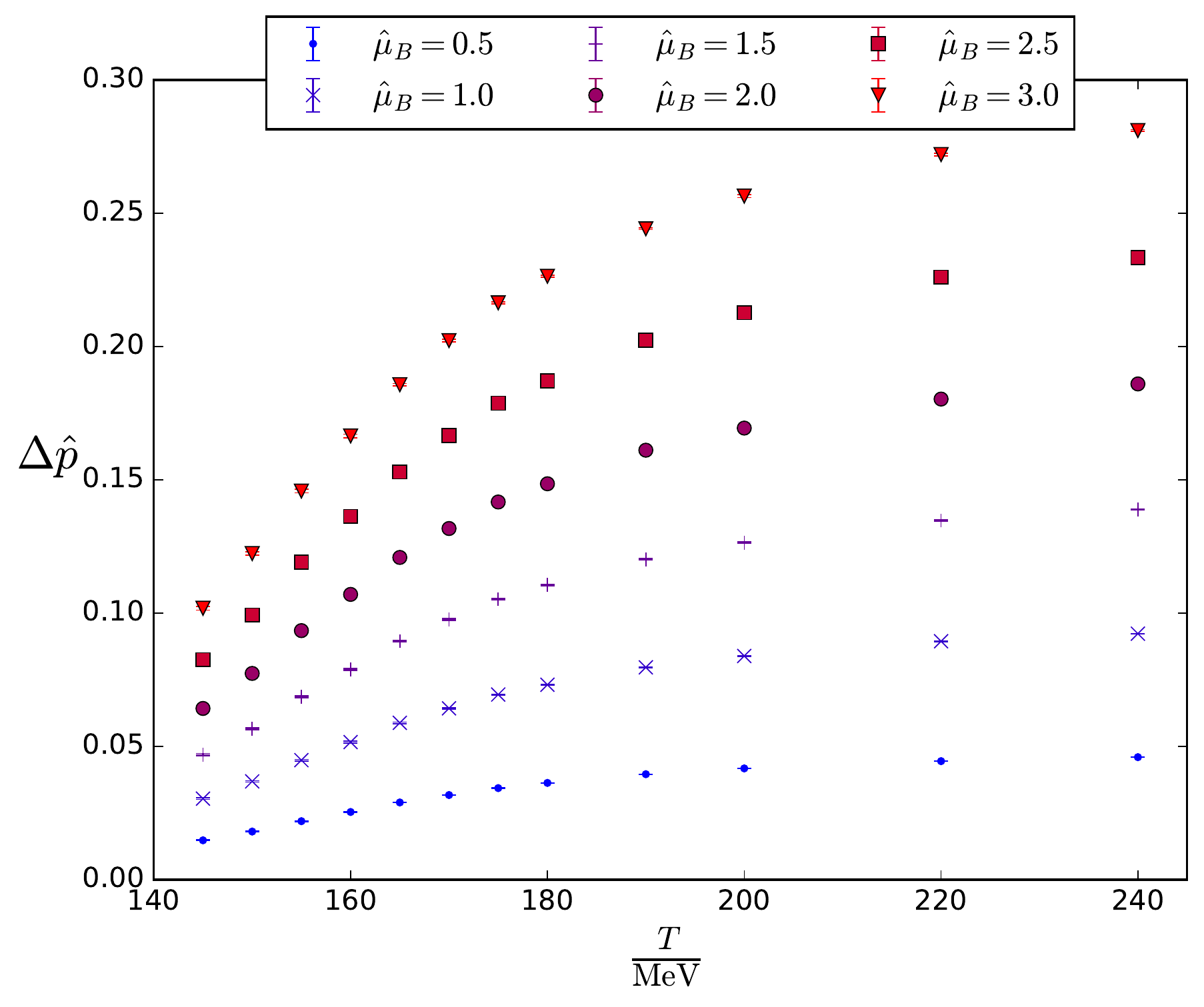}
  \caption{
      \vspace{-0.02cm}
      Left panel: The pressure difference between the zero and finite chemical potential theory calculated with several different reweighting methods
      as a function of the quark chemical potential at $T=160$~MeV. Results at finite isospin chemical potential are shown as purple points. Results reweighted from the 
      phase quenched ensemble to finite baryon density with the two methods discussed in the text are shown by red and blue points. Results from reweighting 
      from zero chemical potential are shown in grey. Right panel: The pressure difference between zero and non-zero baryochemical potentials, calculated with phase reweighting
      as a function of the temperature, for different fixed values of $\hat{\mu}$.
      \vspace{-0.5cm}
  }
  \label{fig:reweighting_methods}
\end{figure*}

\paragraph{Observables}
The grand canonical partition function of lattice QCD is given schematically by:
\begin{equation}
    Z(T,\mu) = \int \Dmeas U\, \det {\textrm{M}}(U,\mu)  e^{-S_g(U)}\,,
\end{equation}
where $S_g(U)$ is the gauge action, $\det \rm{M}$ is the quark determinant, implicitly 
including all flavors, as well as staggered rooting, $\mu$ collectively denotes the
chemical potentials of all quark flavors and $U$ are the link variables. 
We work with the quark chemical potentials 
satisfying $\mu_q \equiv \mu_u=\mu_d$ and $\mu_s=0$.
We concentrate on 
the pressure as 
a function of the temperature and 
dimensionless chemical 
potential $\hat{\mu}_B \equiv \mu_B /T = 3 \mu_q / T \equiv 3 \hat{\mu}_q$
, given by
\begin{equation}
    \hat{p}(T,\hat{\mu}_B) \equiv \frac{p(T, \hat{\mu}_B)}{T^4} = \frac{1}{\left( LT \right)^3} \ln Z(T, \hat{\mu}_B)\rm{.} 
\end{equation}
In particular, we study the pressure difference between zero 
and non-zero chemical potentials:
\begin{equation}
    \Delta \hat{p} = \frac{\Delta p}{T^4} \equiv \hat{p}(T,\hat{\mu}_B) - \hat{p}(T,0)\rm{.}
\end{equation}
We also compute the light quark density:
\begin{equation}
    \begin{aligned}
        \hat{n}_L \left( T, \hat{\mu}_B \right) 
        &\equiv \f{{\rm{d}} \hat{p}}{{\rm{d}} \hat{\mu}_B} = \frac{1}{3 \left( L T \right)^3} \left( \f{\partial
        \ln Z(T,\hat{\mu}_B)}{\partial \hat{\mu}_q} \right)_{\mu_s=0}\rm{.}
    \end{aligned}
\end{equation}
The integral of $\hat{n}_L$ over $\hat{\mu}_B$ is $\Delta \hat{p}$. 
We calculate the equation of state with several different methods.

\paragraph{Reweighting from $\mu_B=0$} 
In what is arguably the simplest reweighting scheme, simulations
are performed at $\mu_B = 0$ and $\Delta \hat{p}$ is reconstructed via~\cite{Barbour:1997ej}:
\begin{equation}
    \label{eq:Glasgow}
    \Delta\hat{p}(T,\hat{\mu}_B) = \frac{1}{\left( LT \right)^3} \ln \left\langle \frac{\det \rm{M}(\hat{\mu}_B)}{\det \rm{M}(0)} \right\rangle_{\hat{\mu}_B=0} \rm{.}
\end{equation}
While Eq.~\eqref{eq:Glasgow} is exact with infinite statistics, the tails 
of the distribution of the weights $\frac{\det {\rm{M}}(\hat{\mu    }_B)}{\det {\rm{M}}(0)}$ are heavy and so hard to sample correctly (overlap problem), 
and it is therefore hard to judge the reliability of results obtained with this method.
It was proposed that the
overlap problem can be mitigated by 
reweighting in more parameters~\cite{Fodor:2001au,Fodor:2001pe,Fodor:2004nz}.
However, even with multi-parameter reweighting, the 
overlap problem remains the main bottleneck of the method (at least around the transition line, where this question was studied in Ref. ~\cite{Giordano:2020uvk}).
We include results from (one-parameter) reweighting from $\mu_B=0$ for 
completeness, and also because one of the resummation strategies we are going 
to test can be regarded as a truncation of this reweighting scheme.

\paragraph{Phase reweighting}

A way to avoid heavy-tailed distributions in the weights is to simulate a theory where 
these can only take values 
in a compact domain. Two examples are sign 
reweighting~\cite{deForcrand:2002pa,Alexandru:2005ix,Giordano:2020roi,Borsanyi:2021hbk}
and phase reweighting~\cite{Fodor:2007vv,Endrodi:2018zda,Borsanyi:2021hbk}. 
Here we use the latter.
Here in the simulated theory 
one replaces the quark determinant with its 
absolute value.
The simulated ensemble is called the phase quenched ensemble,
corresponding to a finite isospin chemical potential, i.e., $\mu_u=-\mu_d$. 
The pressure then schematically reads: 
\begin{equation}
    \hat{p}_I (T,\hat{\mu}_q) = \frac{1}{\left(LT\right)^3} \log \int \mathcal{D} U \left| \det {\rm{M}} \right| e^{-S_g}\rm{.}
\end{equation}
It is also given by the integral of the isospin density:
\begin{equation}
    \hat{n}_I (T, \hat{\mu}_q)  \equiv \left( \frac{\partial \hat{p}_I }{\partial \hat{\mu_q}} \right)_{\mu_s=0}
\end{equation}
from which the pressure at finite $\mu_B$
is obtained as: 
\begin{equation}
    \label{eq:pB_minus_pI}
    \hat{p}(T,\hat{\mu}_B=3\hat{\mu}_q) - \hat{p}_I(T,\hat{\mu}_q) = \frac{1}{\left( LT \right) ^3} \ln \left\langle e^{i \theta} \right\rangle_{PQ} \rm{,}
\end{equation}
where $e^{i \theta} = \frac{\det {\rm{M}}(\hat{\mu}_B)}{\left| \det {\rm{M}}(\hat{\mu}_B) \right|}$ 
is the complex phase factor of the fermion determinant and $\left\langle \dots \right\rangle_{PQ}$ means taking an expectation
value in the phase quenched theory. One finally
arrives at:
\begin{equation}
    \label{eq:pB_from_PQweights}
    \Delta \hat{p} = \int_{0}^{\hat{\mu}_B/3} \hat{n}_I (\hat{\mu}_q, T) {\rm{d}} \hat{\mu}_q 
    +
    \frac{1}{\left( LT \right)^3} \ln \left\langle e^{i \theta} \right\rangle_{PQ}
    \rm{.} 
\end{equation}
Alternatively, one can calculate $\hat{n}_L$ directly with reweighting: 
\begin{equation}
    \label{eq:nl_from_PQ}
    \hat{n}_L = \frac{1}{\left( LT \right)^3 \left\langle e^{i \theta} \right\rangle_{PQ}}
    \left\langle e^{i \theta} \frac{\partial}{\partial \hat{\mu}_B} \ln \det \rm{M} \right\rangle_{PQ}
    \rm{,}
\end{equation}
from which the pressure difference can be calculated via integration. 
The two methods are not a priori guaranteed to give the same results 
if the observable in the numerator of Eq.~\eqref{eq:nl_from_PQ}, namely 
$e^{i \theta} \frac{\partial}{\partial \hat{\mu}_B} \ln \det \rm{M}$, has an overlap problem. 
In principle this could happen, as only the pressure 
difference $p(T,\hat{\mu}_B=\hat{\mu}_q) - p_I(T,\hat{\mu}_I=\hat{\mu}_q) \propto \left\langle e^{i \theta} \right\rangle_{PQ}$ in 
Eq.~\eqref{eq:pB_minus_pI} is guaranteed to be free of one - due to the compactness of the observable $e^{i \theta}$.

The phase diagram at finite isospin density 
was calculated in Ref.~\cite{Brandt:2017oyy}. 
The phase quenched ensemble has a pion condensed phase for low $T$
and $\mu_q > m_\pi/2$. 
In this region, the sign problem
is expected to be severe. 
Here we avoid this issue since we concentrate on 
the equation of state of the QGP, and we 
find a mild sign problem: 
$\left\langle e^{i \theta} \right\rangle_{PQ} = \left\langle \cos \theta \right\rangle_{PQ}$ never gets below $0.1$ in
any of our ensembles and is always more than $10 \sigma$ away from zero.

\paragraph{Taylor expansion}
The pressure is expanded in powers of the baryochemical potential:
\begin{equation}
    \hat{p}(T, \hat{\mu}_B) = p_0 \left( T \right) 
    + p_2 \left( T \right) \hat{\mu}_B^2
    + p_4 \left( T \right) \hat{\mu}_B^4
    + \dots\,.
\end{equation}
We calculate the Taylor expansion coefficients in two different ways: by using configurations generated
at $\mu_B = 0$ to calculate them directly, and 
by using simulations at imaginary chemical potentials to obtain
them from a fit.
These are the two procedures used in the literature so far. For a recent example of 
the first method see, e.g., Ref.~\cite{Bazavov:2020bjn},
for the second one Refs.~\cite{DElia:2016jqh,Borsanyi:2018grb}.
More details on the determination of the coefficients are 
given in the section on our numerical results.
\begin{figure}[t]
    \centering
    \includegraphics[width=0.45\textwidth]{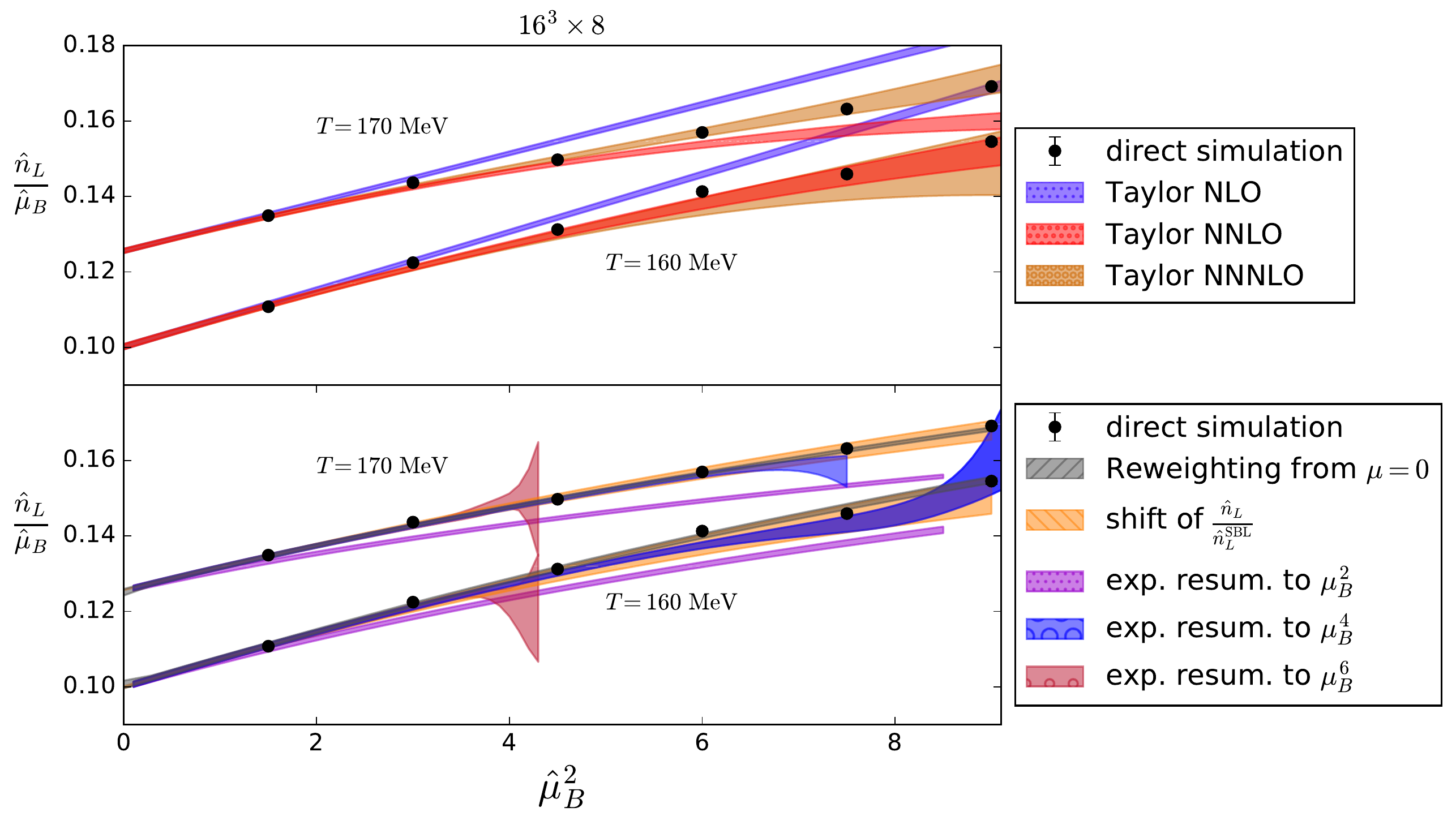}
    \caption{
        \vspace{-0.02cm}
        The direct results for $\hat{n}_L$ at non-zero $\mu_B$ compared with different approximations (or extrapolations):
        the Taylor method to different orders up to $p_8$ 
        and the exponential resummations~\cite{Mondal:2021jxk} to order $N=2, 4$ and $6$, calculated 
        from the ensemble at $\mu_B=0$, as well as the shifting $n_L / n_L^{\rm{SBL}}$ method calculated from imaginary chemical
        potential simulations and reweighting from $\mu_B=0$.
        \vspace{-0.5cm}
    }
    \label{fig:muscan}
\end{figure}

\begin{figure*}[t]
    \centering
    \includegraphics[width=0.45\textwidth]{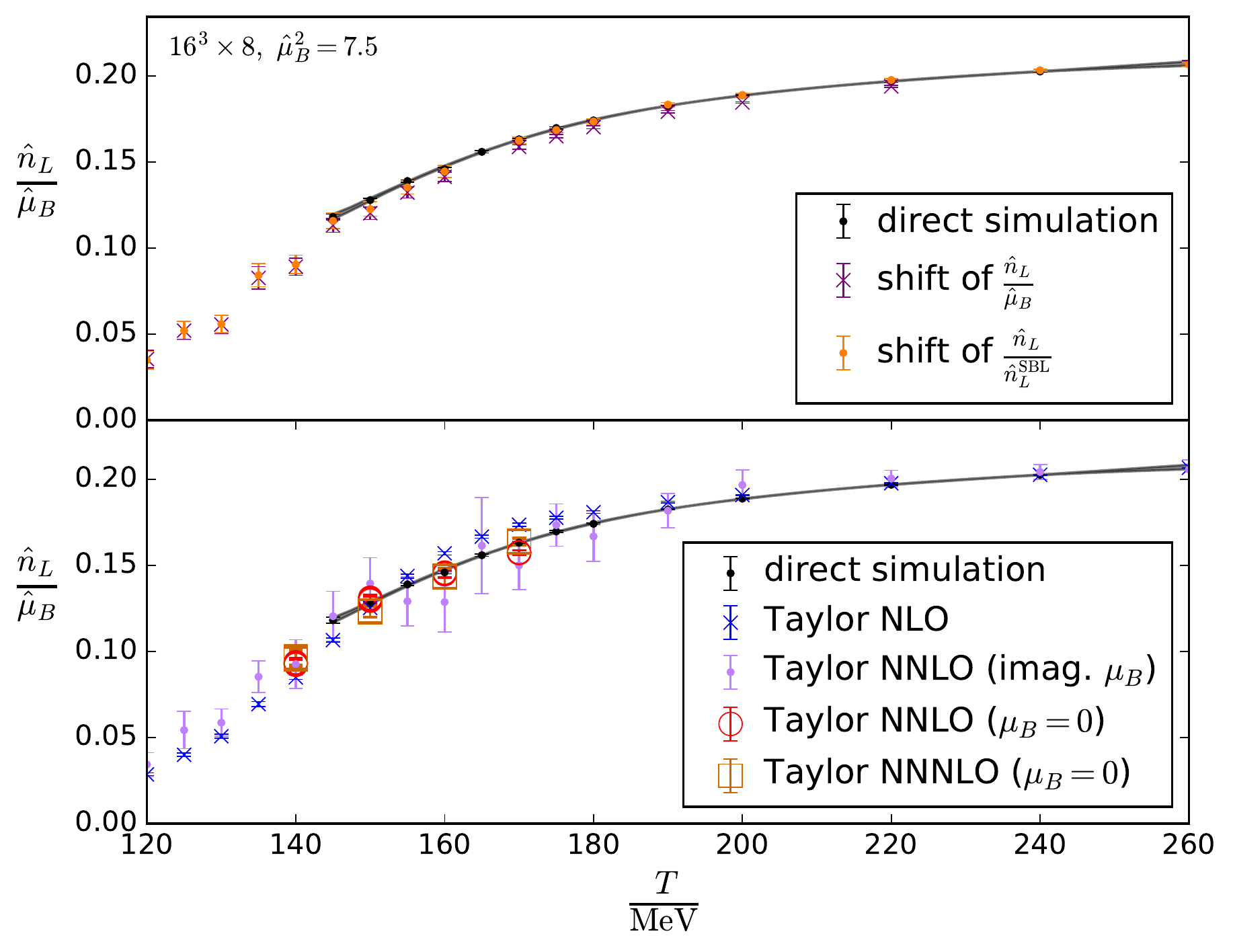}
    \includegraphics[width=0.45\textwidth]{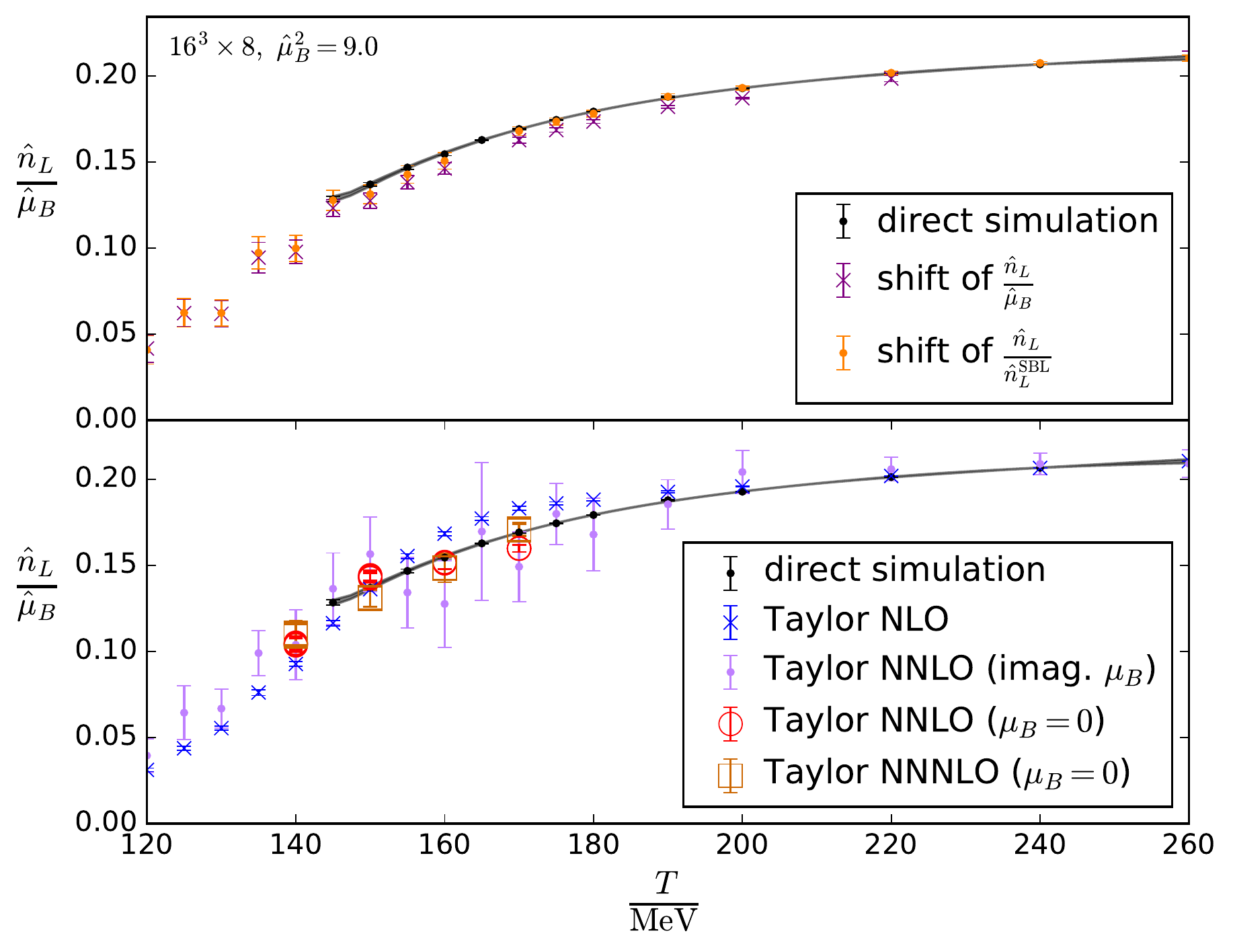}
    \caption{
        \vspace{-0.02cm}
        The density $\hat{n}_L$ as a function of the temperature $T$ for $\hat{\mu}_B = \sqrt{7.5} $ (left) and $\hat{\mu}_B = 3 $ (right) for the ordinary Taylor 
        expansion (bottom) and the resummation schemes based on shifting $\hat{n}_L/\hat{\mu}_B$ and $\hat{n}_L/\hat{n}_L^{\rm{SBL}}$ (top).
        The direct Taylor data from $\mu_B=0$ simulations has smaller errors than 
        the fit to imaginary $\mu_B$ data. This is mainly due to the small 
        volume in our study. For larger volumes, the signal-to-noise ratio of the 
        direct $p_6$ and $p_8$ would be considerably larger. A spline interpolation of the 
        direct results is included to lead the eye.
        \vspace{-0.5cm}
        \label{fig:Tscan}
    }
\end{figure*}

\paragraph{Resummations based on shifting sigmoid functions}
A resummation of the Taylor expansion was introduced in Ref.~\cite{Borsanyi:2021sxv}, defined by the implicit equation
\begin{equation}
    \label{eq:kappaRESUM}
    \frac{\hat{n}_L \left( T, \hat{\mu}_B \right)}{\hat{\mu}_B} = \frac{{\rm{d}} \hat{n}_L}{{\rm{d}} \hat{\mu}_B} \left( T\left( 1 + \kappa_2 \left( T \right) \hat{\mu}_B^2 + \dots \right), 0 \right) \rm{.}
\end{equation}
This is motivated by the empirical observation of 
the existence of an approximate scaling variable $T (1 + \kappa \hat{\mu}_B^2 )$, coming 
from lattice studies at imaginary chemical potential.  These show that certain observables collapse on 
single (sigmoid shaped) curves when plotted against this variable~\cite{Borsanyi:2020fev,Borsanyi:2021sxv}.
Up to $\hat{\mu}_B=1.5$ the existence of an approximate scaling variable 
has also been confirmed directly at a real $\mu_B$ 
by reweighting from the sign quenched 
ensemble~\cite{Borsanyi:2021hbk}. We call this 
the $\frac{\hat{n}_L \left( T, \hat{\mu}_B \right)}{\hat{\mu}_B}$ 
shifting method. 
This is a systematically improvable 
expansion. 
Its validity is not predicated on the existence of this 
approximate scaling variable, only its fast convergence: i.e., 
if an approximate scaling variable exists, we expect the method to
converge fast.

A disadvantage of the previous 
scheme is that it was designed 
to work only near the crossover. 
To make it more suitable for 
high $T$, it was later refined~\cite{Borsanyi:2022qlh} by introducing a Stefan-Boltzmann correction: 
\begin{equation}
    \label{eq:lambdaRESUM}
    \frac{\hat{n}_L \left( T, \hat{\mu}_B \right)}{\hat{n}_L^{\rm{SBL}}\left(\hat{\mu}_B\right)} = \lim_{\hat{\mu}_B' \to 0} \frac{\hat{n}_L \left( T\left( 1 + \lambda_2 \hat{\mu}_B^2 + \dots \right), \hat{\mu}_B' \right) }{\hat{n}_L^{\rm{SBL}}\left( \hat{\mu}_B' \right)} \rm{,}
\end{equation}
where we introduced the Stefan-Boltzmann limit:
\begin{equation}
    \hat{n}_L^{\rm{SBL}}(\hat{\mu}_B) \equiv \lim_{T \to \infty} \hat{n}_L(T, \hat{\mu}_B) \rm{.}
\end{equation}
We call this 
method the $\frac{\hat{n}_L \left( T, \hat{\mu}_B \right)}{\hat{n}_L^{\rm{SBL}}}$ 
shifting method. Both shifting methods can be implemented by fitting the 
$\kappa_n$ or $\lambda_n$ coefficients to data at imaginary $\mu_B$.

\paragraph{Exponential resummation}

A different resummation is 
based on truncating the reweighting from $\mu_B=0$~\cite{Mondal:2021jxk}, by
approximating $\frac{\det \rm{M}(\hat{\mu}_B)}{\det \rm{M}(0)} \simeq \exp \left( \sum_{n=1}^{N} \frac{1}{n!} \mathcal{D}_n \hat{\mu}_B^n \right)$
in Eq.~\eqref{eq:Glasgow}: 
\begin{equation}
    \label{eq:ExpResum}
    \Delta \hat{p}(T,\hat{\mu}_B) \simeq \frac{1}{\left( LT \right)^3} \ln \left\langle \exp \left( \sum_{n=1}^{N} \frac{1}{n!} \mathcal{D}_n \hat{\mu}_B^n  \right) \right\rangle_{\hat{\mu}_B=0} \rm{,}
\end{equation}
where $\mathcal{D}_n \equiv \frac{\partial^n}{\partial \hat{\mu}_B^2} \ln \det \rm{M(\hat{\mu}_B)}$ 
and $N$ is the truncation order.
This method could either be called truncated reweighting or an exponential 
resummation of the Taylor series. A practical 
advantage of this scheme 
is that the coefficients $\mathcal{D}_n$ are needed for the calculation of the 
Taylor expansion coefficients $p_{2n}$ anyway. Thus, this resummation scheme
represents an alternative way to analyze lattice data for the Taylor 
coefficients. The procedure introduced in 
Ref.~\cite{Mondal:2021jxk} had the disadvantage
of using a biased estimator for 
the expectation value in Eq.~\eqref{eq:ExpResum},
by simply exponentiating the stochastic estimators of 
the $\mathcal{D}_n$. 
This bias was studied in Ref.~\cite{Mitra:2022vtf}. 
We remove this bias by calculating the $\mathcal{D}_n$ exactly for each configuration, 
using the reduced matrix formalism~\cite{Hasenfratz:1991ax}. Note that unlike 
the Taylor expansion or the resummations based on shifting sigmoids, this scheme 
is not defined in terms of thermodynamic quantities, but by 
an approximation of the integrand of the path integral. Thus, the convergence
properties of the scheme are not guaranteed to be governed by the 
analytic properties of the free energy.

\paragraph{Lattice setup and numerical results}
We used a tree-level Symanzik improved gauge
action, and two steps of stout smearing~\cite{Morningstar:2003gk} with a
smearing parameter $\rho=0.15$ on the gauge links entering the quark 
determinant, with physical quark masses, using the kaon decay constant $f_K$ for scale
setting (see Ref.~\cite{Aoki:2009sc} for details). 
We apply a tree-level improvement factor to the pressure, that makes 
the $N_\tau = 8$ 2stout results already close to the continuum~\cite{Borsanyi:2010cj}.
This corresponds to division by $1.28$ at infinite volume,
which is the correction we apply for 
$\hat{n}_L$ and $\Delta \hat{p}$. 

We study $16^3\times 8$ lattices at various temperatures in the 
range $145$~MeV $ \leq T\leq 240$~MeV and light-quark
chemical potential $\mu_u=\mu_d=\mu_q=\mu=\mu_B/3$ with a zero strange 
quark chemical potential $\mu_s=0$, corresponding to a strangeness 
chemical potential $\mu_S=\mu_B/3$.  

We simulate
the phase quenched ensemble 
for $\hat{\mu}_B^2 = 1.5,3,4.5,6,7.5$ and $9$.
These simulations are performed
without 
an explicit symmetry breaking term,
commonly used in simulations at a finite isospin 
density~\cite{Kogut:2002zg, Brandt:2017oyy}. 
Instead, we follow the method of Ref.~\cite{Borsanyi:2021hbk}. 
The resulting ensemble corresponds to an isospin chemical 
potential $\mu_I = \mu_u = - \mu_d$. We use these ensembles to 
reweight to a baryochemical potential with $\mu_u = \mu_d = \mu_B/3$.
Determinant ratios
are calculated 
using the reduced matrix formalism~\cite{Hasenfratz:1991ax,Giordano:2019gev}. 
We performed the calculation of $\Delta \hat{p}$ in the two inequivalent ways: the one with no possible overlap problem described by
Eq.~\eqref{eq:pB_from_PQweights}, and by integration of Eq.~\eqref{eq:nl_from_PQ}. We obtain  compatible results for $\Delta \hat{p}$ with the 
two methods, as can be seen in Fig.~\ref{fig:reweighting_methods}~(left) for $T=160$~MeV.

We also simulate at $\mu_B=0$ for four temperatures: $T/{\rm{MeV}}=140,150,160$ and $170$. This is to perform reweighting from $\mu_B=0$, 
to calculate the Taylor expansion coefficients, and to
perform the exponential resummation. 
For all of these, we used the reduced matrix formalism of Ref.~\cite{Hasenfratz:1991ax}, 
so obtaining the Taylor coefficients without the use
of stochastic estimators and obtaining $\mathcal{D}_n$ exactly for each
configuration, and encountering no bias in the
exponential resummation~\cite{Mondal:2021jxk,Mitra:2022vtf}.

In Fig.~\ref{fig:reweighting_methods}~(left)
we also show the results for $\Delta \hat{p}$ from reweighting from $\mu_B=0$.
These results are again in perfect agreement with the two different ways of reweighting from the phase quenched
ensemble. This is true for all temperatures in our study. 

Thus, three different reweighting procedures - including one with no 
possible overlap problem - give identical results. This strongly supports
the validity of our results for the equation of state. The full set of 
results for $\Delta \hat{p}$ from phase reweighting are shown in 
Fig.~\ref{fig:reweighting_methods}~(right). We can safely use these 
results to test extrapolation schemes. 

Without comparison with the phase reweighted results, we 
would have no way to guarantee the 
correctness of reweighting from $\mu_B=0$, due to the uncontrolled
systematics of the overlap problem. This 
problem is also present 
in the Taylor method, since for a finite ensemble the Taylor coefficients
necessarily correspond to the Taylor coefficients of the pressure one would have obtained 
by reweighting from $\mu_B=0$. By first 
showing that reweighting from $\mu_B=0$ works for 
the region $\hat{\mu}_B \leq 3$, we ensure 
that we are truly testing the convergence properties 
of the Taylor series, without being limited by an overlap problem 
in the higher order coefficients.

\paragraph{Comparison with extrapolation schemes}
To implement the resummation
schemes based on shifting sigmoids, we perform simulations
at imaginary 
chemical potentials, for $\operatorname{Im} \hat{\mu}_B \f{16}{\pi}= 0,4,6,7,8,9,10$ and $12$.
We work to order $\kappa_4$ in Eq.~\eqref{eq:kappaRESUM} and to order $\lambda_4$ in Eq.~\eqref{eq:lambdaRESUM}, 
using a simplified version of the analysis 
of Refs.~\cite{Borsanyi:2021sxv,Borsanyi:2022qlh}.
The systematic error includes the fit range in imaginary $\mu_B$, the ansatz in $\mu_B^2$, and the interpolation of the
light quark susceptibility at $\mu_B=0$.
We also demonstrate a more straightforward use of the imaginary chemical potential data by performing a second determination of the Taylor expansion coefficients, fitting
$\f{\hat{n}_L}{\hat{\mu}_B}$ with a polynomial of order $\hat{\mu}_B^6$.
For the fits 
we also include $\f{d^2 \hat{p}}{d \hat{\mu}_B^2}$ 
and $\f{d^4 \hat{p}}{d \hat{\mu}_B^4}$ at $\mu_B=0$ as further datapoints.

We show the comparison of the different reweighting 
schemes with the direct data in Fig.~\ref{fig:Tscan}, as 
a function of the chemical potential at fixed temperatures 
of $T=160$~MeV and $170$~MeV, 
and in Fig.~\ref{fig:muscan}, as 
a function of $T$ at a fixed 
$\hat{\mu}_B = \sqrt{7.5}$ 
and $3$ - the two largest  
values where we have direct data. 
The Taylor expansion at 
next-to-leading order - $\mathcal{O}(\mu_B^4)$ in the pressure - is 
not consistent with the direct data, systematically 
underestimating $n_L$ below $150$~MeV, and systematically overestimating 
it above $150$~MeV.
This is due to a peak in $p_4(T)$ slightly above the crossover temperature. 
Including the next term in the expansion, with 
the coefficient $p_6(T)$, the Taylor method agrees with the direct 
data up to $\hat{\mu}_B=3$ at $T=160$~MeV and up to $\hat{\mu}_B \approx 1.2$ at 
$T=170$~MeV. Including the next-to-next-to-next-to-leading order term $p_8(T)$, 
the expansion agrees with the direct data at all studied 
temperatures up to $\hat{\mu}_B = 3$.  

In contrast, the exponential resummation 
scheme shows bad convergence properties from $\hat{\mu}_B^2 \approx 4$ for all
temperatures. While the $N=2$ truncation of the scheme remains 
close to the direct results 
in the entire range, 
the higher orders make the agreement better only below this value, but not above.

At large $T$, the method 
based on shifting $\hat{n}_L / \hat{n}_L^{\rm{SBL}}$ outperforms 
the method of shifting $\hat{n}_L/\hat{\mu}_B$. This is not 
surprising, as the Stefan-Boltzmann correction was introduced in 
Ref.~\cite{Borsanyi:2022qlh} as a way to improve the convergence 
properties of the scheme at high $T$.

In summary, we can say that both
the Taylor expansion to order $\hat{\mu}_B^8$ and 
the resummation based on shifting $\hat{n}_L / \hat{n}_L^{\rm{SBL}}$ to 
order $\lambda_4$ accurately describe the direct data for the equation of 
state 
in the range $0 \leq \hat{\mu}_B \leq 3$ - which 
includes the entire range of the RHIC Beam 
Energy Scan. Note the faster convergence of the 
resummed expansion, as the calculation of 
the coefficient $\lambda_4$ only requires the 
determination of the Taylor coefficients up to order $\hat{\mu}_B^6$.
On the other hand, the 
shifting $\hat{n}_L / \hat{\mu}_B$ method at order $\kappa_4$ 
has a slight systematic discrepancy with the 
direct data at large $T$, and exponential 
resummation shows bad convergence 
properties in $N$ above $\hat{\mu}_B^2 \approx 4$.

\paragraph{Outlook}
We judged the reliability of different approximation schemes by 
comparing them with direct non-zero chemical potential results in the range
$0 \leq \hat{\mu}_B \leq 3$.
While this gives a practical 
answer to the question of which 
approximation one can trust, a theoretical understanding 
of the reasons would also be welcome. 
For the schemes defined purely in terms of thermodynamic quantities, 
such as the Taylor expansion
or the resummations based on shifting sigmoids, this 
requires knowledge of the 
position of partition function (Lee-Yang) zeros in the complex $\mu_B$ plane~\cite{Giordano:2019slo,Giordano:2019gev,Mukherjee:2021tyg,Dimopoulos:2021vrk,Bollweg:2022rps}.
The exponential resummation scheme, instead, is not defined 
in terms of thermodynamic quantities, 
but comes rather from manipulating the integrand of 
the path integral for the partition function. Understanding its convergence region
might also require better understanding of the 
nuances of the path integral, in addition
to the thermodynamic singularities. 
We speculate that the limited convergence region has to do with quark determinant zeros. 
In fact, the sum appearing as the argument of the exponential in Eq.~\eqref{eq:ExpResum} approximates the effective action 
of the quarks in a fixed gauge field background, with a radius of convergence determined by the determinant zeros, which 
correspond to logarithmic divergences of the effective action. These 
are not simply related to the Lee-Yang zeros of the 
partition function, and may provide stronger limitations 
on the convergence of the expansion.

An obvious challenge is to 
extend the range of validity of the methods studied in this paper 
to lower $T$ and higher $\hat{\mu}_B$, so that the transition 
line~\cite{Bonati:2018nut,HotQCD:2018pds,Borsanyi:2020fev,Pasztor:2020dur,Haque:2020eyj}, 
and the location of the conjectured critical 
endpoint~\cite{Fukushima:2008wg,Kovacs:2016juc,Isserstedt:2019pgx,Gao:2020fbl,Bernhardt:2021iql}
can be studied with first principle lattice calculations.
Of course, the continuum and infinite volume limits will 
also have to be taken eventually.

\paragraph{Acknowledgements}
The project was supported by the BMBF Grant No.~05P21PXFCA.
This work was also supported by the Hungarian National Research,
Development and Innovation Office, NKFIH grant KKP126769.
A.P. is supported by the J. Bolyai Research
Scholarship of the Hungarian Academy of Sciences and by the \'UNKP-21-5 New
National Excellence Program of the Ministry for Innovation and Technology.
The authors gratefully acknowledge the Gauss Centre for Supercomputing
e.V. (www.gauss-centre.eu) for funding this project by providing computing
time on the GCS Supercomputers Jureca/Juwels at Juelich Supercomputer Centre, HAWK at
H\"ochstleistungsrechenzentrum Stuttgart and SuperMUC at Leibniz
Supercomputing Centre.
\newpage


\begin{thebibliography}{69}%
\makeatletter
\providecommand \@ifxundefined [1]{%
 \@ifx{#1\undefined}
}%
\providecommand \@ifnum [1]{%
 \ifnum #1\expandafter \@firstoftwo
 \else \expandafter \@secondoftwo
 \fi
}%
\providecommand \@ifx [1]{%
 \ifx #1\expandafter \@firstoftwo
 \else \expandafter \@secondoftwo
 \fi
}%
\providecommand \natexlab [1]{#1}%
\providecommand \enquote  [1]{``#1''}%
\providecommand \bibnamefont  [1]{#1}%
\providecommand \bibfnamefont [1]{#1}%
\providecommand \citenamefont [1]{#1}%
\providecommand \href@noop [0]{\@secondoftwo}%
\providecommand \href [0]{\begingroup \@sanitize@url \@href}%
\providecommand \@href[1]{\@@startlink{#1}\@@href}%
\providecommand \@@href[1]{\endgroup#1\@@endlink}%
\providecommand \@sanitize@url [0]{\catcode `\\12\catcode `\$12\catcode
  `\&12\catcode `\#12\catcode `\^12\catcode `\_12\catcode `\%12\relax}%
\providecommand \@@startlink[1]{}%
\providecommand \@@endlink[0]{}%
\providecommand \url  [0]{\begingroup\@sanitize@url \@url }%
\providecommand \@url [1]{\endgroup\@href {#1}{\urlprefix }}%
\providecommand \urlprefix  [0]{URL }%
\providecommand \Eprint [0]{\href }%
\providecommand \doibase [0]{https://doi.org/}%
\providecommand \selectlanguage [0]{\@gobble}%
\providecommand \bibinfo  [0]{\@secondoftwo}%
\providecommand \bibfield  [0]{\@secondoftwo}%
\providecommand \translation [1]{[#1]}%
\providecommand \BibitemOpen [0]{}%
\providecommand \bibitemStop [0]{}%
\providecommand \bibitemNoStop [0]{.\EOS\space}%
\providecommand \EOS [0]{\spacefactor3000\relax}%
\providecommand \BibitemShut  [1]{\csname bibitem#1\endcsname}%
\let\auto@bib@innerbib\@empty
\bibitem [{\citenamefont {Montvay}\ and\ \citenamefont
  {M{\"u}nster}(1997)}]{Montvay:1994cy}%
  \BibitemOpen
  \bibfield  {author} {\bibinfo {author} {\bibfnamefont {I.}~\bibnamefont
  {Montvay}}\ and\ \bibinfo {author} {\bibfnamefont {G.}~\bibnamefont
  {M{\"u}nster}},\ }\href {https://doi.org/10.1017/CBO9780511470783} {\emph
  {\bibinfo {title} {{Quantum fields on a lattice}}}},\ Cambridge Monographs on
  Mathematical Physics\ (\bibinfo  {publisher} {Cambridge University Press},\
  \bibinfo {year} {1997})\BibitemShut {NoStop}%
\bibitem [{\citenamefont {Aoki}\ \emph
  {et~al.}(2006{\natexlab{a}})\citenamefont {Aoki}, \citenamefont {Endr{\H
  o}di}, \citenamefont {Fodor}, \citenamefont {Katz},\ and\ \citenamefont
  {Szab{\'o}}}]{Aoki:2006we}%
  \BibitemOpen
  \bibfield  {author} {\bibinfo {author} {\bibfnamefont {Y.}~\bibnamefont
  {Aoki}}, \bibinfo {author} {\bibfnamefont {G.}~\bibnamefont {Endr{\H o}di}},
  \bibinfo {author} {\bibfnamefont {Z.}~\bibnamefont {Fodor}}, \bibinfo
  {author} {\bibfnamefont {S.~D.}\ \bibnamefont {Katz}},\ and\ \bibinfo
  {author} {\bibfnamefont {K.~K.}\ \bibnamefont {Szab{\'o}}},\ }\href
  {https://doi.org/10.1038/nature05120} {\bibfield  {journal} {\bibinfo
  {journal} {Nature}\ }\textbf {\bibinfo {volume} {443}},\ \bibinfo {pages}
  {675} (\bibinfo {year} {2006}{\natexlab{a}})},\ \Eprint
  {https://arxiv.org/abs/hep-lat/0611014} {arXiv:hep-lat/0611014} \BibitemShut
  {NoStop}%
\bibitem [{\citenamefont {Bors{\' a}nyi}\ \emph {et~al.}(2010)\citenamefont
  {Bors{\' a}nyi}, \citenamefont {Fodor}, \citenamefont {Hoelbling},
  \citenamefont {Katz}, \citenamefont {Krieg}, \citenamefont {Ratti},\ and\
  \citenamefont {Szab{\'o}}}]{Borsanyi:2010bp}%
  \BibitemOpen
  \bibfield  {author} {\bibinfo {author} {\bibfnamefont {S.}~\bibnamefont
  {Bors{\' a}nyi}}, \bibinfo {author} {\bibfnamefont {Z.}~\bibnamefont
  {Fodor}}, \bibinfo {author} {\bibfnamefont {C.}~\bibnamefont {Hoelbling}},
  \bibinfo {author} {\bibfnamefont {S.~D.}\ \bibnamefont {Katz}}, \bibinfo
  {author} {\bibfnamefont {S.}~\bibnamefont {Krieg}}, \bibinfo {author}
  {\bibfnamefont {C.}~\bibnamefont {Ratti}},\ and\ \bibinfo {author}
  {\bibfnamefont {K.~K.}\ \bibnamefont {Szab{\'o}}} (\bibinfo {collaboration}
  {Wuppertal-Budapest}),\ }\href {https://doi.org/10.1007/JHEP09(2010)073}
  {\bibfield  {journal} {\bibinfo  {journal} {JHEP}\ }\textbf {\bibinfo
  {volume} {09}},\ \bibinfo {pages} {073}},\ \Eprint
  {https://arxiv.org/abs/1005.3508} {arXiv:1005.3508 [hep-lat]} \BibitemShut
  {NoStop}%
\bibitem [{\citenamefont {Bazavov}\ \emph {et~al.}(2012)\citenamefont {Bazavov}
  \emph {et~al.}}]{Bazavov:2011nk}%
  \BibitemOpen
  \bibfield  {author} {\bibinfo {author} {\bibfnamefont {A.}~\bibnamefont
  {Bazavov}} \emph {et~al.},\ }\href
  {https://doi.org/10.1103/PhysRevD.85.054503} {\bibfield  {journal} {\bibinfo
  {journal} {Phys. Rev. D}\ }\textbf {\bibinfo {volume} {85}},\ \bibinfo
  {pages} {054503} (\bibinfo {year} {2012})},\ \Eprint
  {https://arxiv.org/abs/1111.1710} {arXiv:1111.1710 [hep-lat]} \BibitemShut
  {NoStop}%
\bibitem [{\citenamefont {Bors{\'a}nyi}\ \emph {et~al.}(2010)\citenamefont
  {Bors{\'a}nyi}, \citenamefont {Endr{\H o}di}, \citenamefont {Fodor},
  \citenamefont {Jakov{\' a}c}, \citenamefont {Katz}, \citenamefont {Krieg},
  \citenamefont {Ratti},\ and\ \citenamefont {Szab{\'o}}}]{Borsanyi:2010cj}%
  \BibitemOpen
  \bibfield  {author} {\bibinfo {author} {\bibfnamefont {S.}~\bibnamefont
  {Bors{\'a}nyi}}, \bibinfo {author} {\bibfnamefont {G.}~\bibnamefont {Endr{\H
  o}di}}, \bibinfo {author} {\bibfnamefont {Z.}~\bibnamefont {Fodor}}, \bibinfo
  {author} {\bibfnamefont {A.}~\bibnamefont {Jakov{\' a}c}}, \bibinfo {author}
  {\bibfnamefont {S.~D.}\ \bibnamefont {Katz}}, \bibinfo {author}
  {\bibfnamefont {S.}~\bibnamefont {Krieg}}, \bibinfo {author} {\bibfnamefont
  {C.}~\bibnamefont {Ratti}},\ and\ \bibinfo {author} {\bibfnamefont {K.~K.}\
  \bibnamefont {Szab{\'o}}},\ }\href {https://doi.org/10.1007/JHEP11(2010)077}
  {\bibfield  {journal} {\bibinfo  {journal} {JHEP}\ }\textbf {\bibinfo
  {volume} {11}},\ \bibinfo {pages} {077}},\ \Eprint
  {https://arxiv.org/abs/1007.2580} {arXiv:1007.2580 [hep-lat]} \BibitemShut
  {NoStop}%
\bibitem [{\citenamefont {Bazavov}\ \emph {et~al.}(2014)\citenamefont {Bazavov}
  \emph {et~al.}}]{Bazavov:2014pvz}%
  \BibitemOpen
  \bibfield  {author} {\bibinfo {author} {\bibfnamefont {A.}~\bibnamefont
  {Bazavov}} \emph {et~al.} (\bibinfo {collaboration} {HotQCD}),\ }\href
  {https://doi.org/10.1103/PhysRevD.90.094503} {\bibfield  {journal} {\bibinfo
  {journal} {Phys. Rev. D}\ }\textbf {\bibinfo {volume} {90}},\ \bibinfo
  {pages} {094503} (\bibinfo {year} {2014})},\ \Eprint
  {https://arxiv.org/abs/1407.6387} {arXiv:1407.6387 [hep-lat]} \BibitemShut
  {NoStop}%
\bibitem [{\citenamefont {Gavai}\ and\ \citenamefont
  {Gupta}(2003)}]{Gavai:2003mf}%
  \BibitemOpen
  \bibfield  {author} {\bibinfo {author} {\bibfnamefont {R.~V.}\ \bibnamefont
  {Gavai}}\ and\ \bibinfo {author} {\bibfnamefont {S.}~\bibnamefont {Gupta}},\
  }\href {https://doi.org/10.1103/PhysRevD.68.034506} {\bibfield  {journal}
  {\bibinfo  {journal} {Phys. Rev. D}\ }\textbf {\bibinfo {volume} {68}},\
  \bibinfo {pages} {034506} (\bibinfo {year} {2003})},\ \Eprint
  {https://arxiv.org/abs/hep-lat/0303013} {arXiv:hep-lat/0303013} \BibitemShut
  {NoStop}%
\bibitem [{\citenamefont {Allton}\ \emph {et~al.}(2005)\citenamefont {Allton},
  \citenamefont {Doring}, \citenamefont {Ejiri}, \citenamefont {Hands},
  \citenamefont {Kaczmarek}, \citenamefont {Karsch}, \citenamefont {Laermann},\
  and\ \citenamefont {Redlich}}]{Allton:2005gk}%
  \BibitemOpen
  \bibfield  {author} {\bibinfo {author} {\bibfnamefont {C.~R.}\ \bibnamefont
  {Allton}}, \bibinfo {author} {\bibfnamefont {M.}~\bibnamefont {Doring}},
  \bibinfo {author} {\bibfnamefont {S.}~\bibnamefont {Ejiri}}, \bibinfo
  {author} {\bibfnamefont {S.~J.}\ \bibnamefont {Hands}}, \bibinfo {author}
  {\bibfnamefont {O.}~\bibnamefont {Kaczmarek}}, \bibinfo {author}
  {\bibfnamefont {F.}~\bibnamefont {Karsch}}, \bibinfo {author} {\bibfnamefont
  {E.}~\bibnamefont {Laermann}},\ and\ \bibinfo {author} {\bibfnamefont
  {K.}~\bibnamefont {Redlich}},\ }\href
  {https://doi.org/10.1103/PhysRevD.71.054508} {\bibfield  {journal} {\bibinfo
  {journal} {Phys. Rev. D}\ }\textbf {\bibinfo {volume} {71}},\ \bibinfo
  {pages} {054508} (\bibinfo {year} {2005})},\ \Eprint
  {https://arxiv.org/abs/hep-lat/0501030} {arXiv:hep-lat/0501030 [hep-lat]}
  \BibitemShut {NoStop}%
\bibitem [{\citenamefont {Basak}\ \emph {et~al.}(2008)\citenamefont {Basak}
  \emph {et~al.}}]{MILC:2008reg}%
  \BibitemOpen
  \bibfield  {author} {\bibinfo {author} {\bibfnamefont {S.}~\bibnamefont
  {Basak}} \emph {et~al.} (\bibinfo {collaboration} {MILC}),\ }\href
  {https://doi.org/10.22323/1.066.0171} {\bibfield  {journal} {\bibinfo
  {journal} {PoS}\ }\textbf {\bibinfo {volume} {LATTICE2008}},\ \bibinfo
  {pages} {171} (\bibinfo {year} {2008})},\ \Eprint
  {https://arxiv.org/abs/0910.0276} {arXiv:0910.0276 [hep-lat]} \BibitemShut
  {NoStop}%
\bibitem [{\citenamefont {Bors{\'a}nyi}\ \emph
  {et~al.}(2012{\natexlab{a}})\citenamefont {Bors{\'a}nyi}, \citenamefont
  {Fodor}, \citenamefont {Katz}, \citenamefont {Krieg}, \citenamefont {Ratti},\
  and\ \citenamefont {Szab{\'o}}}]{Borsanyi:2011sw}%
  \BibitemOpen
  \bibfield  {author} {\bibinfo {author} {\bibfnamefont {S.}~\bibnamefont
  {Bors{\'a}nyi}}, \bibinfo {author} {\bibfnamefont {Z.}~\bibnamefont {Fodor}},
  \bibinfo {author} {\bibfnamefont {S.~D.}\ \bibnamefont {Katz}}, \bibinfo
  {author} {\bibfnamefont {S.}~\bibnamefont {Krieg}}, \bibinfo {author}
  {\bibfnamefont {C.}~\bibnamefont {Ratti}},\ and\ \bibinfo {author}
  {\bibfnamefont {K.}~\bibnamefont {Szab{\'o}}},\ }\href
  {https://doi.org/10.1007/JHEP01(2012)138} {\bibfield  {journal} {\bibinfo
  {journal} {JHEP}\ }\textbf {\bibinfo {volume} {01}},\ \bibinfo {pages}
  {138}},\ \Eprint {https://arxiv.org/abs/1112.4416} {arXiv:1112.4416
  [hep-lat]} \BibitemShut {NoStop}%
\bibitem [{\citenamefont {Bors{\'a}nyi}\ \emph
  {et~al.}(2012{\natexlab{b}})\citenamefont {Bors{\'a}nyi}, \citenamefont
  {Endr{\H o}di}, \citenamefont {Fodor}, \citenamefont {Katz}, \citenamefont
  {Krieg} \emph {et~al.}}]{Borsanyi:2012cr}%
  \BibitemOpen
  \bibfield  {author} {\bibinfo {author} {\bibfnamefont {S.}~\bibnamefont
  {Bors{\'a}nyi}}, \bibinfo {author} {\bibfnamefont {G.}~\bibnamefont {Endr{\H
  o}di}}, \bibinfo {author} {\bibfnamefont {Z.}~\bibnamefont {Fodor}}, \bibinfo
  {author} {\bibfnamefont {S.}~\bibnamefont {Katz}}, \bibinfo {author}
  {\bibfnamefont {S.}~\bibnamefont {Krieg}}, \emph {et~al.},\ }\href
  {https://doi.org/10.1007/JHEP08(2012)053} {\bibfield  {journal} {\bibinfo
  {journal} {JHEP}\ }\textbf {\bibinfo {volume} {08}},\ \bibinfo {pages}
  {053}},\ \Eprint {https://arxiv.org/abs/1204.6710} {arXiv:1204.6710
  [hep-lat]} \BibitemShut {NoStop}%
\bibitem [{\citenamefont {Bellwied}\ \emph
  {et~al.}(2015{\natexlab{a}})\citenamefont {Bellwied}, \citenamefont
  {Bors{\'a}nyi}, \citenamefont {Fodor}, \citenamefont {Katz}, \citenamefont
  {P{\'a}sztor}, \citenamefont {Ratti},\ and\ \citenamefont
  {Szab{\'o}}}]{Bellwied:2015lba}%
  \BibitemOpen
  \bibfield  {author} {\bibinfo {author} {\bibfnamefont {R.}~\bibnamefont
  {Bellwied}}, \bibinfo {author} {\bibfnamefont {S.}~\bibnamefont
  {Bors{\'a}nyi}}, \bibinfo {author} {\bibfnamefont {Z.}~\bibnamefont {Fodor}},
  \bibinfo {author} {\bibfnamefont {S.~D.}\ \bibnamefont {Katz}}, \bibinfo
  {author} {\bibfnamefont {A.}~\bibnamefont {P{\'a}sztor}}, \bibinfo {author}
  {\bibfnamefont {C.}~\bibnamefont {Ratti}},\ and\ \bibinfo {author}
  {\bibfnamefont {K.~K.}\ \bibnamefont {Szab{\'o}}},\ }\href
  {https://doi.org/10.1103/PhysRevD.92.114505} {\bibfield  {journal} {\bibinfo
  {journal} {Phys. Rev. D}\ }\textbf {\bibinfo {volume} {92}},\ \bibinfo
  {pages} {114505} (\bibinfo {year} {2015}{\natexlab{a}})},\ \Eprint
  {https://arxiv.org/abs/1507.04627} {arXiv:1507.04627 [hep-lat]} \BibitemShut
  {NoStop}%
\bibitem [{\citenamefont {Ding}\ \emph {et~al.}(2015)\citenamefont {Ding},
  \citenamefont {Mukherjee}, \citenamefont {Ohno}, \citenamefont {Petreczky},\
  and\ \citenamefont {Schadler}}]{Ding:2015fca}%
  \BibitemOpen
  \bibfield  {author} {\bibinfo {author} {\bibfnamefont {H.~T.}\ \bibnamefont
  {Ding}}, \bibinfo {author} {\bibfnamefont {S.}~\bibnamefont {Mukherjee}},
  \bibinfo {author} {\bibfnamefont {H.}~\bibnamefont {Ohno}}, \bibinfo {author}
  {\bibfnamefont {P.}~\bibnamefont {Petreczky}},\ and\ \bibinfo {author}
  {\bibfnamefont {H.~P.}\ \bibnamefont {Schadler}},\ }\href
  {https://doi.org/10.1103/PhysRevD.92.074043} {\bibfield  {journal} {\bibinfo
  {journal} {Phys. Rev. D}\ }\textbf {\bibinfo {volume} {92}},\ \bibinfo
  {pages} {074043} (\bibinfo {year} {2015})},\ \Eprint
  {https://arxiv.org/abs/1507.06637} {arXiv:1507.06637 [hep-lat]} \BibitemShut
  {NoStop}%
\bibitem [{\citenamefont {Bazavov}\ \emph {et~al.}(2017)\citenamefont {Bazavov}
  \emph {et~al.}}]{Bazavov:2017dus}%
  \BibitemOpen
  \bibfield  {author} {\bibinfo {author} {\bibfnamefont {A.}~\bibnamefont
  {Bazavov}} \emph {et~al.},\ }\href
  {https://doi.org/10.1103/PhysRevD.95.054504} {\bibfield  {journal} {\bibinfo
  {journal} {Phys. Rev. D}\ }\textbf {\bibinfo {volume} {95}},\ \bibinfo
  {pages} {054504} (\bibinfo {year} {2017})},\ \Eprint
  {https://arxiv.org/abs/1701.04325} {arXiv:1701.04325 [hep-lat]} \BibitemShut
  {NoStop}%
\bibitem [{\citenamefont {Bazavov}\ \emph {et~al.}(2019)\citenamefont {Bazavov}
  \emph {et~al.}}]{HotQCD:2018pds}%
  \BibitemOpen
  \bibfield  {author} {\bibinfo {author} {\bibfnamefont {A.}~\bibnamefont
  {Bazavov}} \emph {et~al.} (\bibinfo {collaboration} {HotQCD}),\ }\href
  {https://doi.org/10.1016/j.physletb.2019.05.013} {\bibfield  {journal}
  {\bibinfo  {journal} {Phys. Lett. B}\ }\textbf {\bibinfo {volume} {795}},\
  \bibinfo {pages} {15} (\bibinfo {year} {2019})},\ \Eprint
  {https://arxiv.org/abs/1812.08235} {arXiv:1812.08235 [hep-lat]} \BibitemShut
  {NoStop}%
\bibitem [{\citenamefont {Giordano}\ and\ \citenamefont
  {P\'asztor}(2019)}]{Giordano:2019slo}%
  \BibitemOpen
  \bibfield  {author} {\bibinfo {author} {\bibfnamefont {M.}~\bibnamefont
  {Giordano}}\ and\ \bibinfo {author} {\bibfnamefont {A.}~\bibnamefont
  {P\'asztor}},\ }\href {https://doi.org/10.1103/PhysRevD.99.114510} {\bibfield
   {journal} {\bibinfo  {journal} {Phys. Rev. D}\ }\textbf {\bibinfo {volume}
  {99}},\ \bibinfo {pages} {114510} (\bibinfo {year} {2019})},\ \Eprint
  {https://arxiv.org/abs/1904.01974} {arXiv:1904.01974 [hep-lat]} \BibitemShut
  {NoStop}%
\bibitem [{\citenamefont {Bazavov}\ \emph {et~al.}(2020)\citenamefont {Bazavov}
  \emph {et~al.}}]{Bazavov:2020bjn}%
  \BibitemOpen
  \bibfield  {author} {\bibinfo {author} {\bibfnamefont {A.}~\bibnamefont
  {Bazavov}} \emph {et~al.},\ }\href
  {https://doi.org/10.1103/PhysRevD.101.074502} {\bibfield  {journal} {\bibinfo
   {journal} {Phys. Rev. D}\ }\textbf {\bibinfo {volume} {101}},\ \bibinfo
  {pages} {074502} (\bibinfo {year} {2020})},\ \Eprint
  {https://arxiv.org/abs/2001.08530} {arXiv:2001.08530 [hep-lat]} \BibitemShut
  {NoStop}%
\bibitem [{\citenamefont {de~Forcrand}\ and\ \citenamefont
  {Philipsen}(2002)}]{deForcrand:2002hgr}%
  \BibitemOpen
  \bibfield  {author} {\bibinfo {author} {\bibfnamefont {P.}~\bibnamefont
  {de~Forcrand}}\ and\ \bibinfo {author} {\bibfnamefont {O.}~\bibnamefont
  {Philipsen}},\ }\href {https://doi.org/10.1016/S0550-3213(02)00626-0}
  {\bibfield  {journal} {\bibinfo  {journal} {Nucl. Phys. B}\ }\textbf
  {\bibinfo {volume} {642}},\ \bibinfo {pages} {290} (\bibinfo {year}
  {2002})},\ \Eprint {https://arxiv.org/abs/hep-lat/0205016}
  {arXiv:hep-lat/0205016 [hep-lat]} \BibitemShut {NoStop}%
\bibitem [{\citenamefont {D'Elia}\ and\ \citenamefont
  {Lombardo}(2003)}]{DElia:2002tig}%
  \BibitemOpen
  \bibfield  {author} {\bibinfo {author} {\bibfnamefont {M.}~\bibnamefont
  {D'Elia}}\ and\ \bibinfo {author} {\bibfnamefont {M.~P.}\ \bibnamefont
  {Lombardo}},\ }\href {https://doi.org/10.1103/PhysRevD.67.014505} {\bibfield
  {journal} {\bibinfo  {journal} {Phys. Rev. D}\ }\textbf {\bibinfo {volume}
  {67}},\ \bibinfo {pages} {014505} (\bibinfo {year} {2003})},\ \Eprint
  {https://arxiv.org/abs/hep-lat/0209146} {arXiv:hep-lat/0209146 [hep-lat]}
  \BibitemShut {NoStop}%
\bibitem [{\citenamefont {D'Elia}\ and\ \citenamefont
  {Sanfilippo}(2009)}]{DElia:2009pdy}%
  \BibitemOpen
  \bibfield  {author} {\bibinfo {author} {\bibfnamefont {M.}~\bibnamefont
  {D'Elia}}\ and\ \bibinfo {author} {\bibfnamefont {F.}~\bibnamefont
  {Sanfilippo}},\ }\href {https://doi.org/10.1103/PhysRevD.80.014502}
  {\bibfield  {journal} {\bibinfo  {journal} {Phys. Rev. D}\ }\textbf {\bibinfo
  {volume} {80}},\ \bibinfo {pages} {014502} (\bibinfo {year} {2009})},\
  \Eprint {https://arxiv.org/abs/0904.1400} {arXiv:0904.1400 [hep-lat]}
  \BibitemShut {NoStop}%
\bibitem [{\citenamefont {Cea}\ \emph {et~al.}(2014)\citenamefont {Cea},
  \citenamefont {Cosmai},\ and\ \citenamefont {Papa}}]{Cea:2014xva}%
  \BibitemOpen
  \bibfield  {author} {\bibinfo {author} {\bibfnamefont {P.}~\bibnamefont
  {Cea}}, \bibinfo {author} {\bibfnamefont {L.}~\bibnamefont {Cosmai}},\ and\
  \bibinfo {author} {\bibfnamefont {A.}~\bibnamefont {Papa}},\ }\href
  {https://doi.org/10.1103/PhysRevD.89.074512} {\bibfield  {journal} {\bibinfo
  {journal} {Phys. Rev. D}\ }\textbf {\bibinfo {volume} {89}},\ \bibinfo
  {pages} {074512} (\bibinfo {year} {2014})},\ \Eprint
  {https://arxiv.org/abs/1403.0821} {arXiv:1403.0821 [hep-lat]} \BibitemShut
  {NoStop}%
\bibitem [{\citenamefont {Bonati}\ \emph {et~al.}(2014)\citenamefont {Bonati},
  \citenamefont {de~Forcrand}, \citenamefont {D'Elia}, \citenamefont
  {Philipsen},\ and\ \citenamefont {Sanfilippo}}]{Bonati:2014kpa}%
  \BibitemOpen
  \bibfield  {author} {\bibinfo {author} {\bibfnamefont {C.}~\bibnamefont
  {Bonati}}, \bibinfo {author} {\bibfnamefont {P.}~\bibnamefont {de~Forcrand}},
  \bibinfo {author} {\bibfnamefont {M.}~\bibnamefont {D'Elia}}, \bibinfo
  {author} {\bibfnamefont {O.}~\bibnamefont {Philipsen}},\ and\ \bibinfo
  {author} {\bibfnamefont {F.}~\bibnamefont {Sanfilippo}},\ }\href
  {https://doi.org/10.1103/PhysRevD.90.074030} {\bibfield  {journal} {\bibinfo
  {journal} {Phys. Rev. D}\ }\textbf {\bibinfo {volume} {90}},\ \bibinfo
  {pages} {074030} (\bibinfo {year} {2014})},\ \Eprint
  {https://arxiv.org/abs/1408.5086} {arXiv:1408.5086 [hep-lat]} \BibitemShut
  {NoStop}%
\bibitem [{\citenamefont {Cea}\ \emph {et~al.}(2016)\citenamefont {Cea},
  \citenamefont {Cosmai},\ and\ \citenamefont {Papa}}]{Cea:2015cya}%
  \BibitemOpen
  \bibfield  {author} {\bibinfo {author} {\bibfnamefont {P.}~\bibnamefont
  {Cea}}, \bibinfo {author} {\bibfnamefont {L.}~\bibnamefont {Cosmai}},\ and\
  \bibinfo {author} {\bibfnamefont {A.}~\bibnamefont {Papa}},\ }\href
  {https://doi.org/10.1103/PhysRevD.93.014507} {\bibfield  {journal} {\bibinfo
  {journal} {Phys. Rev. D}\ }\textbf {\bibinfo {volume} {93}},\ \bibinfo
  {pages} {014507} (\bibinfo {year} {2016})},\ \Eprint
  {https://arxiv.org/abs/1508.07599} {arXiv:1508.07599 [hep-lat]} \BibitemShut
  {NoStop}%
\bibitem [{\citenamefont {Bonati}\ \emph {et~al.}(2015)\citenamefont {Bonati},
  \citenamefont {D'Elia}, \citenamefont {Mariti}, \citenamefont {Mesiti},
  \citenamefont {Negro},\ and\ \citenamefont {Sanfilippo}}]{Bonati:2015bha}%
  \BibitemOpen
  \bibfield  {author} {\bibinfo {author} {\bibfnamefont {C.}~\bibnamefont
  {Bonati}}, \bibinfo {author} {\bibfnamefont {M.}~\bibnamefont {D'Elia}},
  \bibinfo {author} {\bibfnamefont {M.}~\bibnamefont {Mariti}}, \bibinfo
  {author} {\bibfnamefont {M.}~\bibnamefont {Mesiti}}, \bibinfo {author}
  {\bibfnamefont {F.}~\bibnamefont {Negro}},\ and\ \bibinfo {author}
  {\bibfnamefont {F.}~\bibnamefont {Sanfilippo}},\ }\href
  {https://doi.org/10.1103/PhysRevD.92.054503} {\bibfield  {journal} {\bibinfo
  {journal} {Phys. Rev. D}\ }\textbf {\bibinfo {volume} {92}},\ \bibinfo
  {pages} {054503} (\bibinfo {year} {2015})},\ \Eprint
  {https://arxiv.org/abs/1507.03571} {arXiv:1507.03571 [hep-lat]} \BibitemShut
  {NoStop}%
\bibitem [{\citenamefont {Bellwied}\ \emph
  {et~al.}(2015{\natexlab{b}})\citenamefont {Bellwied}, \citenamefont
  {Bors{\'a}nyi}, \citenamefont {Fodor}, \citenamefont {G{\"u}nther},
  \citenamefont {Katz}, \citenamefont {Ratti},\ and\ \citenamefont
  {Szab{\'o}}}]{Bellwied:2015rza}%
  \BibitemOpen
  \bibfield  {author} {\bibinfo {author} {\bibfnamefont {R.}~\bibnamefont
  {Bellwied}}, \bibinfo {author} {\bibfnamefont {S.}~\bibnamefont
  {Bors{\'a}nyi}}, \bibinfo {author} {\bibfnamefont {Z.}~\bibnamefont {Fodor}},
  \bibinfo {author} {\bibfnamefont {J.}~\bibnamefont {G{\"u}nther}}, \bibinfo
  {author} {\bibfnamefont {S.~D.}\ \bibnamefont {Katz}}, \bibinfo {author}
  {\bibfnamefont {C.}~\bibnamefont {Ratti}},\ and\ \bibinfo {author}
  {\bibfnamefont {K.~K.}\ \bibnamefont {Szab{\'o}}},\ }\href
  {https://doi.org/10.1016/j.physletb.2015.11.011} {\bibfield  {journal}
  {\bibinfo  {journal} {Phys. Lett. B}\ }\textbf {\bibinfo {volume} {751}},\
  \bibinfo {pages} {559} (\bibinfo {year} {2015}{\natexlab{b}})},\ \Eprint
  {https://arxiv.org/abs/1507.07510} {arXiv:1507.07510 [hep-lat]} \BibitemShut
  {NoStop}%
\bibitem [{\citenamefont {D'Elia}\ \emph {et~al.}(2017)\citenamefont {D'Elia},
  \citenamefont {Gagliardi},\ and\ \citenamefont {Sanfilippo}}]{DElia:2016jqh}%
  \BibitemOpen
  \bibfield  {author} {\bibinfo {author} {\bibfnamefont {M.}~\bibnamefont
  {D'Elia}}, \bibinfo {author} {\bibfnamefont {G.}~\bibnamefont {Gagliardi}},\
  and\ \bibinfo {author} {\bibfnamefont {F.}~\bibnamefont {Sanfilippo}},\
  }\href {https://doi.org/10.1103/PhysRevD.95.094503} {\bibfield  {journal}
  {\bibinfo  {journal} {Phys. Rev. D}\ }\textbf {\bibinfo {volume} {95}},\
  \bibinfo {pages} {094503} (\bibinfo {year} {2017})},\ \Eprint
  {https://arxiv.org/abs/1611.08285} {arXiv:1611.08285 [hep-lat]} \BibitemShut
  {NoStop}%
\bibitem [{\citenamefont {G{\"u}nther}\ \emph {et~al.}(2017)\citenamefont
  {G{\"u}nther}, \citenamefont {Bellwied}, \citenamefont {Bors{\'a}nyi},
  \citenamefont {Fodor}, \citenamefont {Katz}, \citenamefont {P{\'a}sztor},
  \citenamefont {Ratti},\ and\ \citenamefont {Szab{\'o}}}]{Gunther:2016vcp}%
  \BibitemOpen
  \bibfield  {author} {\bibinfo {author} {\bibfnamefont {J.~N.}\ \bibnamefont
  {G{\"u}nther}}, \bibinfo {author} {\bibfnamefont {R.}~\bibnamefont
  {Bellwied}}, \bibinfo {author} {\bibfnamefont {S.}~\bibnamefont
  {Bors{\'a}nyi}}, \bibinfo {author} {\bibfnamefont {Z.}~\bibnamefont {Fodor}},
  \bibinfo {author} {\bibfnamefont {S.~D.}\ \bibnamefont {Katz}}, \bibinfo
  {author} {\bibfnamefont {A.}~\bibnamefont {P{\'a}sztor}}, \bibinfo {author}
  {\bibfnamefont {C.}~\bibnamefont {Ratti}},\ and\ \bibinfo {author}
  {\bibfnamefont {K.~K.}\ \bibnamefont {Szab{\'o}}},\ }\bibfield  {booktitle}
  {\emph {\bibinfo {booktitle} {{Proceedings, 26th International Conference on
  Ultra-relativistic Nucleus-Nucleus Collisions (Quark Matter 2017): Chicago,
  Illinois, USA, February 5-11, 2017}}},\ }\href
  {https://doi.org/10.1016/j.nuclphysa.2017.05.044} {\bibfield  {journal}
  {\bibinfo  {journal} {Nucl. Phys.}\ }\textbf {\bibinfo {volume} {A967}},\
  \bibinfo {pages} {720} (\bibinfo {year} {2017})},\ \Eprint
  {https://arxiv.org/abs/1607.02493} {arXiv:1607.02493 [hep-lat]} \BibitemShut
  {NoStop}%
\bibitem [{\citenamefont {Alba}\ \emph {et~al.}(2017)\citenamefont {Alba} \emph
  {et~al.}}]{Alba:2017mqu}%
  \BibitemOpen
  \bibfield  {author} {\bibinfo {author} {\bibfnamefont {P.}~\bibnamefont
  {Alba}} \emph {et~al.},\ }\href {https://doi.org/10.1103/PhysRevD.96.034517}
  {\bibfield  {journal} {\bibinfo  {journal} {Phys. Rev. D}\ }\textbf {\bibinfo
  {volume} {96}},\ \bibinfo {pages} {034517} (\bibinfo {year} {2017})},\
  \Eprint {https://arxiv.org/abs/1702.01113} {arXiv:1702.01113 [hep-lat]}
  \BibitemShut {NoStop}%
\bibitem [{\citenamefont {Vovchenko}\ \emph {et~al.}(2017)\citenamefont
  {Vovchenko}, \citenamefont {P{\'a}sztor}, \citenamefont {Fodor},
  \citenamefont {Katz},\ and\ \citenamefont {Stoecker}}]{Vovchenko:2017xad}%
  \BibitemOpen
  \bibfield  {author} {\bibinfo {author} {\bibfnamefont {V.}~\bibnamefont
  {Vovchenko}}, \bibinfo {author} {\bibfnamefont {A.}~\bibnamefont
  {P{\'a}sztor}}, \bibinfo {author} {\bibfnamefont {Z.}~\bibnamefont {Fodor}},
  \bibinfo {author} {\bibfnamefont {S.~D.}\ \bibnamefont {Katz}},\ and\
  \bibinfo {author} {\bibfnamefont {H.}~\bibnamefont {Stoecker}},\ }\href
  {https://doi.org/10.1016/j.physletb.2017.10.042} {\bibfield  {journal}
  {\bibinfo  {journal} {Phys. Lett. B}\ }\textbf {\bibinfo {volume} {775}},\
  \bibinfo {pages} {71} (\bibinfo {year} {2017})},\ \Eprint
  {https://arxiv.org/abs/1708.02852} {arXiv:1708.02852 [hep-ph]} \BibitemShut
  {NoStop}%
\bibitem [{\citenamefont {Bonati}\ \emph {et~al.}(2018)\citenamefont {Bonati},
  \citenamefont {D'Elia}, \citenamefont {Negro}, \citenamefont {Sanfilippo},\
  and\ \citenamefont {Zambello}}]{Bonati:2018nut}%
  \BibitemOpen
  \bibfield  {author} {\bibinfo {author} {\bibfnamefont {C.}~\bibnamefont
  {Bonati}}, \bibinfo {author} {\bibfnamefont {M.}~\bibnamefont {D'Elia}},
  \bibinfo {author} {\bibfnamefont {F.}~\bibnamefont {Negro}}, \bibinfo
  {author} {\bibfnamefont {F.}~\bibnamefont {Sanfilippo}},\ and\ \bibinfo
  {author} {\bibfnamefont {K.}~\bibnamefont {Zambello}},\ }\href
  {https://doi.org/10.1103/PhysRevD.98.054510} {\bibfield  {journal} {\bibinfo
  {journal} {Phys. Rev. D}\ }\textbf {\bibinfo {volume} {98}},\ \bibinfo
  {pages} {054510} (\bibinfo {year} {2018})},\ \Eprint
  {https://arxiv.org/abs/1805.02960} {arXiv:1805.02960 [hep-lat]} \BibitemShut
  {NoStop}%
\bibitem [{\citenamefont {Bors{\'a}nyi}\ \emph {et~al.}(2018)\citenamefont
  {Bors{\'a}nyi}, \citenamefont {Fodor}, \citenamefont {G{\"u}nther},
  \citenamefont {Katz}, \citenamefont {Szab{\'o}}, \citenamefont {P{\'a}sztor},
  \citenamefont {Portillo},\ and\ \citenamefont {Ratti}}]{Borsanyi:2018grb}%
  \BibitemOpen
  \bibfield  {author} {\bibinfo {author} {\bibfnamefont {S.}~\bibnamefont
  {Bors{\'a}nyi}}, \bibinfo {author} {\bibfnamefont {Z.}~\bibnamefont {Fodor}},
  \bibinfo {author} {\bibfnamefont {J.~N.}\ \bibnamefont {G{\"u}nther}},
  \bibinfo {author} {\bibfnamefont {S.~K.}\ \bibnamefont {Katz}}, \bibinfo
  {author} {\bibfnamefont {K.~K.}\ \bibnamefont {Szab{\'o}}}, \bibinfo {author}
  {\bibfnamefont {A.}~\bibnamefont {P{\'a}sztor}}, \bibinfo {author}
  {\bibfnamefont {I.}~\bibnamefont {Portillo}},\ and\ \bibinfo {author}
  {\bibfnamefont {C.}~\bibnamefont {Ratti}},\ }\href
  {https://doi.org/10.1007/JHEP10(2018)205} {\bibfield  {journal} {\bibinfo
  {journal} {JHEP}\ }\textbf {\bibinfo {volume} {10}},\ \bibinfo {pages}
  {205}},\ \Eprint {https://arxiv.org/abs/1805.04445} {arXiv:1805.04445
  [hep-lat]} \BibitemShut {NoStop}%
\bibitem [{\citenamefont {Bellwied}\ \emph {et~al.}(2020)\citenamefont
  {Bellwied}, \citenamefont {Bors{\'a}nyi}, \citenamefont {Fodor},
  \citenamefont {G{\"u}nther}, \citenamefont {Noronha-Hostler}, \citenamefont
  {Parotto}, \citenamefont {P{\'a}sztor}, \citenamefont {Ratti},\ and\
  \citenamefont {Stafford}}]{Bellwied:2019pxh}%
  \BibitemOpen
  \bibfield  {author} {\bibinfo {author} {\bibfnamefont {R.}~\bibnamefont
  {Bellwied}}, \bibinfo {author} {\bibfnamefont {S.}~\bibnamefont
  {Bors{\'a}nyi}}, \bibinfo {author} {\bibfnamefont {Z.}~\bibnamefont {Fodor}},
  \bibinfo {author} {\bibfnamefont {J.~N.}\ \bibnamefont {G{\"u}nther}},
  \bibinfo {author} {\bibfnamefont {J.}~\bibnamefont {Noronha-Hostler}},
  \bibinfo {author} {\bibfnamefont {P.}~\bibnamefont {Parotto}}, \bibinfo
  {author} {\bibfnamefont {A.}~\bibnamefont {P{\'a}sztor}}, \bibinfo {author}
  {\bibfnamefont {C.}~\bibnamefont {Ratti}},\ and\ \bibinfo {author}
  {\bibfnamefont {J.~M.}\ \bibnamefont {Stafford}},\ }\href
  {https://doi.org/10.1103/PhysRevD.101.034506} {\bibfield  {journal} {\bibinfo
   {journal} {Phys. Rev. D}\ }\textbf {\bibinfo {volume} {101}},\ \bibinfo
  {pages} {034506} (\bibinfo {year} {2020})},\ \Eprint
  {https://arxiv.org/abs/1910.14592} {arXiv:1910.14592 [hep-lat]} \BibitemShut
  {NoStop}%
\bibitem [{\citenamefont {Bors{\' a}nyi}\ \emph {et~al.}(2020)\citenamefont
  {Bors{\' a}nyi}, \citenamefont {Fodor}, \citenamefont {G{\"u}nther},
  \citenamefont {Kara}, \citenamefont {Katz}, \citenamefont {Parotto},
  \citenamefont {P{\'a}sztor}, \citenamefont {Ratti},\ and\ \citenamefont
  {Szab{\'o}}}]{Borsanyi:2020fev}%
  \BibitemOpen
  \bibfield  {author} {\bibinfo {author} {\bibfnamefont {S.}~\bibnamefont
  {Bors{\' a}nyi}}, \bibinfo {author} {\bibfnamefont {Z.}~\bibnamefont
  {Fodor}}, \bibinfo {author} {\bibfnamefont {J.~N.}\ \bibnamefont
  {G{\"u}nther}}, \bibinfo {author} {\bibfnamefont {R.}~\bibnamefont {Kara}},
  \bibinfo {author} {\bibfnamefont {S.~D.}\ \bibnamefont {Katz}}, \bibinfo
  {author} {\bibfnamefont {P.}~\bibnamefont {Parotto}}, \bibinfo {author}
  {\bibfnamefont {A.}~\bibnamefont {P{\'a}sztor}}, \bibinfo {author}
  {\bibfnamefont {C.}~\bibnamefont {Ratti}},\ and\ \bibinfo {author}
  {\bibfnamefont {K.~K.}\ \bibnamefont {Szab{\'o}}},\ }\href
  {https://doi.org/10.1103/PhysRevLett.125.052001} {\bibfield  {journal}
  {\bibinfo  {journal} {Phys. Rev. Lett.}\ }\textbf {\bibinfo {volume} {125}},\
  \bibinfo {pages} {052001} (\bibinfo {year} {2020})},\ \Eprint
  {https://arxiv.org/abs/2002.02821} {arXiv:2002.02821 [hep-lat]} \BibitemShut
  {NoStop}%
\bibitem [{\citenamefont {Bors\'anyi}\ \emph {et~al.}(2021)\citenamefont
  {Bors\'anyi}, \citenamefont {Fodor}, \citenamefont {G{\"u}nther},
  \citenamefont {Kara}, \citenamefont {Katz}, \citenamefont {Parotto},
  \citenamefont {P\'asztor}, \citenamefont {Ratti},\ and\ \citenamefont
  {Szab\'o}}]{Borsanyi:2021sxv}%
  \BibitemOpen
  \bibfield  {author} {\bibinfo {author} {\bibfnamefont {S.}~\bibnamefont
  {Bors\'anyi}}, \bibinfo {author} {\bibfnamefont {Z.}~\bibnamefont {Fodor}},
  \bibinfo {author} {\bibfnamefont {J.~N.}\ \bibnamefont {G{\"u}nther}},
  \bibinfo {author} {\bibfnamefont {R.}~\bibnamefont {Kara}}, \bibinfo {author}
  {\bibfnamefont {S.~D.}\ \bibnamefont {Katz}}, \bibinfo {author}
  {\bibfnamefont {P.}~\bibnamefont {Parotto}}, \bibinfo {author} {\bibfnamefont
  {A.}~\bibnamefont {P\'asztor}}, \bibinfo {author} {\bibfnamefont
  {C.}~\bibnamefont {Ratti}},\ and\ \bibinfo {author} {\bibfnamefont {K.~K.}\
  \bibnamefont {Szab\'o}},\ }\href
  {https://doi.org/10.1103/PhysRevLett.126.232001} {\bibfield  {journal}
  {\bibinfo  {journal} {Phys. Rev. Lett.}\ }\textbf {\bibinfo {volume} {126}},\
  \bibinfo {pages} {232001} (\bibinfo {year} {2021})},\ \Eprint
  {https://arxiv.org/abs/2102.06660} {arXiv:2102.06660 [hep-lat]} \BibitemShut
  {NoStop}%
\bibitem [{\citenamefont {Mondal}\ \emph {et~al.}(2022)\citenamefont {Mondal},
  \citenamefont {Mukherjee},\ and\ \citenamefont {Hegde}}]{Mondal:2021jxk}%
  \BibitemOpen
  \bibfield  {author} {\bibinfo {author} {\bibfnamefont {S.}~\bibnamefont
  {Mondal}}, \bibinfo {author} {\bibfnamefont {S.}~\bibnamefont {Mukherjee}},\
  and\ \bibinfo {author} {\bibfnamefont {P.}~\bibnamefont {Hegde}},\ }\href
  {https://doi.org/10.1103/PhysRevLett.128.022001} {\bibfield  {journal}
  {\bibinfo  {journal} {Phys. Rev. Lett.}\ }\textbf {\bibinfo {volume} {128}},\
  \bibinfo {pages} {022001} (\bibinfo {year} {2022})},\ \Eprint
  {https://arxiv.org/abs/2106.03165} {arXiv:2106.03165 [hep-lat]} \BibitemShut
  {NoStop}%
\bibitem [{\citenamefont {Mitra}\ \emph {et~al.}(2022)\citenamefont {Mitra},
  \citenamefont {Hegde},\ and\ \citenamefont {Schmidt}}]{Mitra:2022vtf}%
  \BibitemOpen
  \bibfield  {author} {\bibinfo {author} {\bibfnamefont {S.}~\bibnamefont
  {Mitra}}, \bibinfo {author} {\bibfnamefont {P.}~\bibnamefont {Hegde}},\ and\
  \bibinfo {author} {\bibfnamefont {C.}~\bibnamefont {Schmidt}},\ }\href@noop
  {} {\  (\bibinfo {year} {2022})},\ \Eprint {https://arxiv.org/abs/2205.08517}
  {arXiv:2205.08517 [hep-lat]} \BibitemShut {NoStop}%
\bibitem [{\citenamefont {Bollweg}\ \emph {et~al.}(2022)\citenamefont
  {Bollweg}, \citenamefont {Goswami}, \citenamefont {Kaczmarek}, \citenamefont
  {Karsch}, \citenamefont {Mukherjee}, \citenamefont {Petreczky}, \citenamefont
  {Schmidt},\ and\ \citenamefont {Scior}}]{Bollweg:2022rps}%
  \BibitemOpen
  \bibfield  {author} {\bibinfo {author} {\bibfnamefont {D.}~\bibnamefont
  {Bollweg}}, \bibinfo {author} {\bibfnamefont {J.}~\bibnamefont {Goswami}},
  \bibinfo {author} {\bibfnamefont {O.}~\bibnamefont {Kaczmarek}}, \bibinfo
  {author} {\bibfnamefont {F.}~\bibnamefont {Karsch}}, \bibinfo {author}
  {\bibfnamefont {S.}~\bibnamefont {Mukherjee}}, \bibinfo {author}
  {\bibfnamefont {P.}~\bibnamefont {Petreczky}}, \bibinfo {author}
  {\bibfnamefont {C.}~\bibnamefont {Schmidt}},\ and\ \bibinfo {author}
  {\bibfnamefont {P.}~\bibnamefont {Scior}} (\bibinfo {collaboration}
  {HotQCD}),\ }\href {https://doi.org/10.1103/PhysRevD.105.074511} {\bibfield
  {journal} {\bibinfo  {journal} {Phys. Rev. D}\ }\textbf {\bibinfo {volume}
  {105}},\ \bibinfo {pages} {074511} (\bibinfo {year} {2022})},\ \Eprint
  {https://arxiv.org/abs/2202.09184} {arXiv:2202.09184 [hep-lat]} \BibitemShut
  {NoStop}%
\bibitem [{\citenamefont {Borsanyi}\ \emph
  {et~al.}(2022{\natexlab{a}})\citenamefont {Borsanyi}, \citenamefont {Fodor},
  \citenamefont {Guenther}, \citenamefont {Kara}, \citenamefont {Parotto},
  \citenamefont {Pasztor}, \citenamefont {Ratti},\ and\ \citenamefont
  {Szabo}}]{Borsanyi:2022qlh}%
  \BibitemOpen
  \bibfield  {author} {\bibinfo {author} {\bibfnamefont {S.}~\bibnamefont
  {Borsanyi}}, \bibinfo {author} {\bibfnamefont {Z.}~\bibnamefont {Fodor}},
  \bibinfo {author} {\bibfnamefont {J.~N.}\ \bibnamefont {Guenther}}, \bibinfo
  {author} {\bibfnamefont {R.}~\bibnamefont {Kara}}, \bibinfo {author}
  {\bibfnamefont {P.}~\bibnamefont {Parotto}}, \bibinfo {author} {\bibfnamefont
  {A.}~\bibnamefont {Pasztor}}, \bibinfo {author} {\bibfnamefont
  {C.}~\bibnamefont {Ratti}},\ and\ \bibinfo {author} {\bibfnamefont {K.~K.}\
  \bibnamefont {Szabo}},\ }\href {https://doi.org/10.1103/PhysRevD.105.114504}
  {\bibfield  {journal} {\bibinfo  {journal} {Phys. Rev. D}\ }\textbf {\bibinfo
  {volume} {105}},\ \bibinfo {pages} {114504} (\bibinfo {year}
  {2022}{\natexlab{a}})},\ \Eprint {https://arxiv.org/abs/2202.05574}
  {arXiv:2202.05574 [hep-lat]} \BibitemShut {NoStop}%
\bibitem [{\citenamefont {Giordano}\ \emph
  {et~al.}(2020{\natexlab{a}})\citenamefont {Giordano}, \citenamefont
  {Kap{\'a}s}, \citenamefont {Katz}, \citenamefont {N{\'o}gr{\'a}di},\ and\
  \citenamefont {P{\'a}sztor}}]{Giordano:2020roi}%
  \BibitemOpen
  \bibfield  {author} {\bibinfo {author} {\bibfnamefont {M.}~\bibnamefont
  {Giordano}}, \bibinfo {author} {\bibfnamefont {K.}~\bibnamefont {Kap{\'a}s}},
  \bibinfo {author} {\bibfnamefont {S.~D.}\ \bibnamefont {Katz}}, \bibinfo
  {author} {\bibfnamefont {D.}~\bibnamefont {N{\'o}gr{\'a}di}},\ and\ \bibinfo
  {author} {\bibfnamefont {A.}~\bibnamefont {P{\'a}sztor}},\ }\href
  {https://doi.org/10.1007/JHEP05(2020)088} {\bibfield  {journal} {\bibinfo
  {journal} {JHEP}\ }\textbf {\bibinfo {volume} {05}},\ \bibinfo {pages}
  {088}},\ \Eprint {https://arxiv.org/abs/2004.10800} {arXiv:2004.10800
  [hep-lat]} \BibitemShut {NoStop}%
\bibitem [{\citenamefont {Borsanyi}\ \emph
  {et~al.}(2022{\natexlab{b}})\citenamefont {Borsanyi}, \citenamefont {Fodor},
  \citenamefont {Giordano}, \citenamefont {Katz}, \citenamefont {Nogradi},
  \citenamefont {Pasztor},\ and\ \citenamefont {Wong}}]{Borsanyi:2021hbk}%
  \BibitemOpen
  \bibfield  {author} {\bibinfo {author} {\bibfnamefont {S.}~\bibnamefont
  {Borsanyi}}, \bibinfo {author} {\bibfnamefont {Z.}~\bibnamefont {Fodor}},
  \bibinfo {author} {\bibfnamefont {M.}~\bibnamefont {Giordano}}, \bibinfo
  {author} {\bibfnamefont {S.~D.}\ \bibnamefont {Katz}}, \bibinfo {author}
  {\bibfnamefont {D.}~\bibnamefont {Nogradi}}, \bibinfo {author} {\bibfnamefont
  {A.}~\bibnamefont {Pasztor}},\ and\ \bibinfo {author} {\bibfnamefont {C.~H.}\
  \bibnamefont {Wong}},\ }\href {https://doi.org/10.1103/PhysRevD.105.L051506}
  {\bibfield  {journal} {\bibinfo  {journal} {Phys. Rev. D}\ }\textbf {\bibinfo
  {volume} {105}},\ \bibinfo {pages} {L051506} (\bibinfo {year}
  {2022}{\natexlab{b}})},\ \Eprint {https://arxiv.org/abs/2108.09213}
  {arXiv:2108.09213 [hep-lat]} \BibitemShut {NoStop}%
\bibitem [{\citenamefont {Aoki}\ \emph
  {et~al.}(2006{\natexlab{b}})\citenamefont {Aoki}, \citenamefont {Fodor},
  \citenamefont {Katz},\ and\ \citenamefont {Szab{\'o}}}]{Aoki:2006br}%
  \BibitemOpen
  \bibfield  {author} {\bibinfo {author} {\bibfnamefont {Y.}~\bibnamefont
  {Aoki}}, \bibinfo {author} {\bibfnamefont {Z.}~\bibnamefont {Fodor}},
  \bibinfo {author} {\bibfnamefont {S.~D.}\ \bibnamefont {Katz}},\ and\
  \bibinfo {author} {\bibfnamefont {K.~K.}\ \bibnamefont {Szab{\'o}}},\ }\href
  {https://doi.org/10.1016/j.physletb.2006.10.021} {\bibfield  {journal}
  {\bibinfo  {journal} {Phys. Lett. B}\ }\textbf {\bibinfo {volume} {643}},\
  \bibinfo {pages} {46} (\bibinfo {year} {2006}{\natexlab{b}})},\ \Eprint
  {https://arxiv.org/abs/hep-lat/0609068} {arXiv:hep-lat/0609068} \BibitemShut
  {NoStop}%
\bibitem [{\citenamefont {Bali}\ \emph
  {et~al.}(2012{\natexlab{a}})\citenamefont {Bali}, \citenamefont {Bruckmann},
  \citenamefont {Endr{\H o}di}, \citenamefont {Fodor}, \citenamefont {Katz},
  \citenamefont {Krieg}, \citenamefont {Sch{\"a}fer},\ and\ \citenamefont
  {Szab{\'o}}}]{Bali:2011qj}%
  \BibitemOpen
  \bibfield  {author} {\bibinfo {author} {\bibfnamefont {G.~S.}\ \bibnamefont
  {Bali}}, \bibinfo {author} {\bibfnamefont {F.}~\bibnamefont {Bruckmann}},
  \bibinfo {author} {\bibfnamefont {G.}~\bibnamefont {Endr{\H o}di}}, \bibinfo
  {author} {\bibfnamefont {Z.}~\bibnamefont {Fodor}}, \bibinfo {author}
  {\bibfnamefont {S.~D.}\ \bibnamefont {Katz}}, \bibinfo {author}
  {\bibfnamefont {S.}~\bibnamefont {Krieg}}, \bibinfo {author} {\bibfnamefont
  {A.}~\bibnamefont {Sch{\"a}fer}},\ and\ \bibinfo {author} {\bibfnamefont
  {K.~K.}\ \bibnamefont {Szab{\'o}}},\ }\href
  {https://doi.org/10.1007/JHEP02(2012)044} {\bibfield  {journal} {\bibinfo
  {journal} {JHEP}\ }\textbf {\bibinfo {volume} {02}},\ \bibinfo {pages}
  {044}},\ \Eprint {https://arxiv.org/abs/1111.4956} {arXiv:1111.4956
  [hep-lat]} \BibitemShut {NoStop}%
\bibitem [{\citenamefont {Bali}\ \emph
  {et~al.}(2012{\natexlab{b}})\citenamefont {Bali}, \citenamefont {Bruckmann},
  \citenamefont {Endr{\H o}di}, \citenamefont {Fodor}, \citenamefont {Katz},\
  and\ \citenamefont {Sch{\"a}fer}}]{Bali:2012zg}%
  \BibitemOpen
  \bibfield  {author} {\bibinfo {author} {\bibfnamefont {G.~S.}\ \bibnamefont
  {Bali}}, \bibinfo {author} {\bibfnamefont {F.}~\bibnamefont {Bruckmann}},
  \bibinfo {author} {\bibfnamefont {G.}~\bibnamefont {Endr{\H o}di}}, \bibinfo
  {author} {\bibfnamefont {Z.}~\bibnamefont {Fodor}}, \bibinfo {author}
  {\bibfnamefont {S.~D.}\ \bibnamefont {Katz}},\ and\ \bibinfo {author}
  {\bibfnamefont {A.}~\bibnamefont {Sch{\"a}fer}},\ }\href
  {https://doi.org/10.1103/PhysRevD.86.071502} {\bibfield  {journal} {\bibinfo
  {journal} {Phys. Rev. D}\ }\textbf {\bibinfo {volume} {86}},\ \bibinfo
  {pages} {071502} (\bibinfo {year} {2012}{\natexlab{b}})},\ \Eprint
  {https://arxiv.org/abs/1206.4205} {arXiv:1206.4205 [hep-lat]} \BibitemShut
  {NoStop}%
\bibitem [{\citenamefont {Bors\'anyi}\ \emph {et~al.}(2015)\citenamefont
  {Bors\'anyi}, \citenamefont {Fodor}, \citenamefont {Katz}, \citenamefont
  {P\'asztor}, \citenamefont {Szab\'o},\ and\ \citenamefont
  {T\"or\"ok}}]{Borsanyi:2015yka}%
  \BibitemOpen
  \bibfield  {author} {\bibinfo {author} {\bibfnamefont {S.}~\bibnamefont
  {Bors\'anyi}}, \bibinfo {author} {\bibfnamefont {Z.}~\bibnamefont {Fodor}},
  \bibinfo {author} {\bibfnamefont {S.~D.}\ \bibnamefont {Katz}}, \bibinfo
  {author} {\bibfnamefont {A.}~\bibnamefont {P\'asztor}}, \bibinfo {author}
  {\bibfnamefont {K.~K.}\ \bibnamefont {Szab\'o}},\ and\ \bibinfo {author}
  {\bibfnamefont {C.}~\bibnamefont {T\"or\"ok}},\ }\href
  {https://doi.org/10.1007/JHEP04(2015)138} {\bibfield  {journal} {\bibinfo
  {journal} {JHEP}\ }\textbf {\bibinfo {volume} {04}},\ \bibinfo {pages}
  {138}},\ \Eprint {https://arxiv.org/abs/1501.02173} {arXiv:1501.02173
  [hep-lat]} \BibitemShut {NoStop}%
\bibitem [{\citenamefont {Brandt}\ \emph {et~al.}(2018)\citenamefont {Brandt},
  \citenamefont {Endr{\H o}di},\ and\ \citenamefont
  {Schmalzbauer}}]{Brandt:2017oyy}%
  \BibitemOpen
  \bibfield  {author} {\bibinfo {author} {\bibfnamefont {B.~B.}\ \bibnamefont
  {Brandt}}, \bibinfo {author} {\bibfnamefont {G.}~\bibnamefont {Endr{\H
  o}di}},\ and\ \bibinfo {author} {\bibfnamefont {S.}~\bibnamefont
  {Schmalzbauer}},\ }\href {https://doi.org/10.1103/PhysRevD.97.054514}
  {\bibfield  {journal} {\bibinfo  {journal} {Phys. Rev. D}\ }\textbf {\bibinfo
  {volume} {97}},\ \bibinfo {pages} {054514} (\bibinfo {year} {2018})},\
  \Eprint {https://arxiv.org/abs/1712.08190} {arXiv:1712.08190 [hep-lat]}
  \BibitemShut {NoStop}%
\bibitem [{\citenamefont {D'Elia}\ \emph {et~al.}(2019)\citenamefont {D'Elia},
  \citenamefont {Negro}, \citenamefont {Rucci},\ and\ \citenamefont
  {Sanfilippo}}]{DElia:2019iis}%
  \BibitemOpen
  \bibfield  {author} {\bibinfo {author} {\bibfnamefont {M.}~\bibnamefont
  {D'Elia}}, \bibinfo {author} {\bibfnamefont {F.}~\bibnamefont {Negro}},
  \bibinfo {author} {\bibfnamefont {A.}~\bibnamefont {Rucci}},\ and\ \bibinfo
  {author} {\bibfnamefont {F.}~\bibnamefont {Sanfilippo}},\ }\href
  {https://doi.org/10.1103/PhysRevD.100.054504} {\bibfield  {journal} {\bibinfo
   {journal} {Phys. Rev. D}\ }\textbf {\bibinfo {volume} {100}},\ \bibinfo
  {pages} {054504} (\bibinfo {year} {2019})},\ \Eprint
  {https://arxiv.org/abs/1907.09461} {arXiv:1907.09461 [hep-lat]} \BibitemShut
  {NoStop}%
\bibitem [{\citenamefont {Barbour}\ \emph {et~al.}(1998)\citenamefont
  {Barbour}, \citenamefont {Morrison}, \citenamefont {Klepfish}, \citenamefont
  {Kogut},\ and\ \citenamefont {Lombardo}}]{Barbour:1997ej}%
  \BibitemOpen
  \bibfield  {author} {\bibinfo {author} {\bibfnamefont {I.~M.}\ \bibnamefont
  {Barbour}}, \bibinfo {author} {\bibfnamefont {S.~E.}\ \bibnamefont
  {Morrison}}, \bibinfo {author} {\bibfnamefont {E.~G.}\ \bibnamefont
  {Klepfish}}, \bibinfo {author} {\bibfnamefont {J.~B.}\ \bibnamefont
  {Kogut}},\ and\ \bibinfo {author} {\bibfnamefont {M.-P.}\ \bibnamefont
  {Lombardo}},\ }\href {https://doi.org/10.1016/S0920-5632(97)00484-2}
  {\bibfield  {journal} {\bibinfo  {journal} {Nucl. Phys. B Proc. Suppl.}\
  }\textbf {\bibinfo {volume} {60}},\ \bibinfo {pages} {220} (\bibinfo {year}
  {1998})},\ \Eprint {https://arxiv.org/abs/hep-lat/9705042}
  {arXiv:hep-lat/9705042} \BibitemShut {NoStop}%
\bibitem [{\citenamefont {Fodor}\ and\ \citenamefont
  {Katz}(2002{\natexlab{a}})}]{Fodor:2001au}%
  \BibitemOpen
  \bibfield  {author} {\bibinfo {author} {\bibfnamefont {Z.}~\bibnamefont
  {Fodor}}\ and\ \bibinfo {author} {\bibfnamefont {S.~D.}\ \bibnamefont
  {Katz}},\ }\href {https://doi.org/10.1016/S0370-2693(02)01583-6} {\bibfield
  {journal} {\bibinfo  {journal} {Phys. Lett. B}\ }\textbf {\bibinfo {volume}
  {534}},\ \bibinfo {pages} {87} (\bibinfo {year} {2002}{\natexlab{a}})},\
  \Eprint {https://arxiv.org/abs/hep-lat/0104001} {arXiv:hep-lat/0104001}
  \BibitemShut {NoStop}%
\bibitem [{\citenamefont {Fodor}\ and\ \citenamefont
  {Katz}(2002{\natexlab{b}})}]{Fodor:2001pe}%
  \BibitemOpen
  \bibfield  {author} {\bibinfo {author} {\bibfnamefont {Z.}~\bibnamefont
  {Fodor}}\ and\ \bibinfo {author} {\bibfnamefont {S.~D.}\ \bibnamefont
  {Katz}},\ }\href {https://doi.org/10.1088/1126-6708/2002/03/014} {\bibfield
  {journal} {\bibinfo  {journal} {JHEP}\ }\textbf {\bibinfo {volume} {03}},\
  \bibinfo {pages} {014}},\ \Eprint {https://arxiv.org/abs/hep-lat/0106002}
  {arXiv:hep-lat/0106002} \BibitemShut {NoStop}%
\bibitem [{\citenamefont {Fodor}\ and\ \citenamefont
  {Katz}(2004)}]{Fodor:2004nz}%
  \BibitemOpen
  \bibfield  {author} {\bibinfo {author} {\bibfnamefont {Z.}~\bibnamefont
  {Fodor}}\ and\ \bibinfo {author} {\bibfnamefont {S.~D.}\ \bibnamefont
  {Katz}},\ }\href {https://doi.org/10.1088/1126-6708/2004/04/050} {\bibfield
  {journal} {\bibinfo  {journal} {JHEP}\ }\textbf {\bibinfo {volume} {04}},\
  \bibinfo {pages} {050}},\ \Eprint {https://arxiv.org/abs/hep-lat/0402006}
  {arXiv:hep-lat/0402006} \BibitemShut {NoStop}%
\bibitem [{\citenamefont {Giordano}\ \emph
  {et~al.}(2020{\natexlab{b}})\citenamefont {Giordano}, \citenamefont {Kapas},
  \citenamefont {Katz}, \citenamefont {Nogradi},\ and\ \citenamefont
  {Pasztor}}]{Giordano:2020uvk}%
  \BibitemOpen
  \bibfield  {author} {\bibinfo {author} {\bibfnamefont {M.}~\bibnamefont
  {Giordano}}, \bibinfo {author} {\bibfnamefont {K.}~\bibnamefont {Kapas}},
  \bibinfo {author} {\bibfnamefont {S.~D.}\ \bibnamefont {Katz}}, \bibinfo
  {author} {\bibfnamefont {D.}~\bibnamefont {Nogradi}},\ and\ \bibinfo {author}
  {\bibfnamefont {A.}~\bibnamefont {Pasztor}},\ }\href
  {https://doi.org/10.1103/PhysRevD.102.034503} {\bibfield  {journal} {\bibinfo
   {journal} {Phys. Rev. D}\ }\textbf {\bibinfo {volume} {102}},\ \bibinfo
  {pages} {034503} (\bibinfo {year} {2020}{\natexlab{b}})},\ \Eprint
  {https://arxiv.org/abs/2003.04355} {arXiv:2003.04355 [hep-lat]} \BibitemShut
  {NoStop}%
\bibitem [{\citenamefont {de~Forcrand}\ \emph {et~al.}(2003)\citenamefont
  {de~Forcrand}, \citenamefont {Kim},\ and\ \citenamefont
  {Takaishi}}]{deForcrand:2002pa}%
  \BibitemOpen
  \bibfield  {author} {\bibinfo {author} {\bibfnamefont {P.}~\bibnamefont
  {de~Forcrand}}, \bibinfo {author} {\bibfnamefont {S.}~\bibnamefont {Kim}},\
  and\ \bibinfo {author} {\bibfnamefont {T.}~\bibnamefont {Takaishi}},\ }\href
  {https://doi.org/10.1016/S0920-5632(03)80451-6} {\bibfield  {journal}
  {\bibinfo  {journal} {Nucl. Phys. B Proc. Suppl.}\ }\textbf {\bibinfo
  {volume} {119}},\ \bibinfo {pages} {541} (\bibinfo {year} {2003})},\ \Eprint
  {https://arxiv.org/abs/hep-lat/0209126} {arXiv:hep-lat/0209126} \BibitemShut
  {NoStop}%
\bibitem [{\citenamefont {Alexandru}\ \emph {et~al.}(2005)\citenamefont
  {Alexandru}, \citenamefont {Faber}, \citenamefont {Horv{\'a}th},\ and\
  \citenamefont {Liu}}]{Alexandru:2005ix}%
  \BibitemOpen
  \bibfield  {author} {\bibinfo {author} {\bibfnamefont {A.}~\bibnamefont
  {Alexandru}}, \bibinfo {author} {\bibfnamefont {M.}~\bibnamefont {Faber}},
  \bibinfo {author} {\bibfnamefont {I.}~\bibnamefont {Horv{\'a}th}},\ and\
  \bibinfo {author} {\bibfnamefont {K.-F.}\ \bibnamefont {Liu}},\ }\href
  {https://doi.org/10.1103/PhysRevD.72.114513} {\bibfield  {journal} {\bibinfo
  {journal} {Phys. Rev. D}\ }\textbf {\bibinfo {volume} {72}},\ \bibinfo
  {pages} {114513} (\bibinfo {year} {2005})},\ \Eprint
  {https://arxiv.org/abs/hep-lat/0507020} {arXiv:hep-lat/0507020} \BibitemShut
  {NoStop}%
\bibitem [{\citenamefont {Fodor}\ \emph {et~al.}(2007)\citenamefont {Fodor},
  \citenamefont {Katz},\ and\ \citenamefont {Schmidt}}]{Fodor:2007vv}%
  \BibitemOpen
  \bibfield  {author} {\bibinfo {author} {\bibfnamefont {Z.}~\bibnamefont
  {Fodor}}, \bibinfo {author} {\bibfnamefont {S.~D.}\ \bibnamefont {Katz}},\
  and\ \bibinfo {author} {\bibfnamefont {C.}~\bibnamefont {Schmidt}},\ }\href
  {https://doi.org/10.1088/1126-6708/2007/03/121} {\bibfield  {journal}
  {\bibinfo  {journal} {JHEP}\ }\textbf {\bibinfo {volume} {03}},\ \bibinfo
  {pages} {121}},\ \Eprint {https://arxiv.org/abs/hep-lat/0701022}
  {arXiv:hep-lat/0701022} \BibitemShut {NoStop}%
\bibitem [{\citenamefont {Endr{\H o}di}\ \emph {et~al.}(2018)\citenamefont
  {Endr{\H o}di}, \citenamefont {Fodor}, \citenamefont {Katz}, \citenamefont
  {Sexty}, \citenamefont {Szab{\'o}},\ and\ \citenamefont
  {T{\"o}r{\"o}k}}]{Endrodi:2018zda}%
  \BibitemOpen
  \bibfield  {author} {\bibinfo {author} {\bibfnamefont {G.}~\bibnamefont
  {Endr{\H o}di}}, \bibinfo {author} {\bibfnamefont {Z.}~\bibnamefont {Fodor}},
  \bibinfo {author} {\bibfnamefont {S.~D.}\ \bibnamefont {Katz}}, \bibinfo
  {author} {\bibfnamefont {D.}~\bibnamefont {Sexty}}, \bibinfo {author}
  {\bibfnamefont {K.~K.}\ \bibnamefont {Szab{\'o}}},\ and\ \bibinfo {author}
  {\bibfnamefont {C.}~\bibnamefont {T{\"o}r{\"o}k}},\ }\href
  {https://doi.org/10.1103/PhysRevD.98.074508} {\bibfield  {journal} {\bibinfo
  {journal} {Phys. Rev. D}\ }\textbf {\bibinfo {volume} {98}},\ \bibinfo
  {pages} {074508} (\bibinfo {year} {2018})},\ \Eprint
  {https://arxiv.org/abs/1807.08326} {arXiv:1807.08326 [hep-lat]} \BibitemShut
  {NoStop}%
\bibitem [{\citenamefont {Hasenfratz}\ and\ \citenamefont
  {Toussaint}(1992)}]{Hasenfratz:1991ax}%
  \BibitemOpen
  \bibfield  {author} {\bibinfo {author} {\bibfnamefont {A.}~\bibnamefont
  {Hasenfratz}}\ and\ \bibinfo {author} {\bibfnamefont {D.}~\bibnamefont
  {Toussaint}},\ }\href {https://doi.org/10.1016/0550-3213(92)90247-9}
  {\bibfield  {journal} {\bibinfo  {journal} {Nucl. Phys. B}\ }\textbf
  {\bibinfo {volume} {371}},\ \bibinfo {pages} {539} (\bibinfo {year}
  {1992})}\BibitemShut {NoStop}%
\bibitem [{\citenamefont {Morningstar}\ and\ \citenamefont
  {Peardon}(2004)}]{Morningstar:2003gk}%
  \BibitemOpen
  \bibfield  {author} {\bibinfo {author} {\bibfnamefont {C.}~\bibnamefont
  {Morningstar}}\ and\ \bibinfo {author} {\bibfnamefont {M.~J.}\ \bibnamefont
  {Peardon}},\ }\href {https://doi.org/10.1103/PhysRevD.69.054501} {\bibfield
  {journal} {\bibinfo  {journal} {Phys. Rev. D}\ }\textbf {\bibinfo {volume}
  {69}},\ \bibinfo {pages} {054501} (\bibinfo {year} {2004})},\ \Eprint
  {https://arxiv.org/abs/hep-lat/0311018} {arXiv:hep-lat/0311018} \BibitemShut
  {NoStop}%
\bibitem [{\citenamefont {Aoki}\ \emph {et~al.}(2009)\citenamefont {Aoki},
  \citenamefont {Bors{\'a}nyi}, \citenamefont {Durr}, \citenamefont {Fodor},
  \citenamefont {Katz}, \citenamefont {Krieg},\ and\ \citenamefont
  {Szab{\'o}}}]{Aoki:2009sc}%
  \BibitemOpen
  \bibfield  {author} {\bibinfo {author} {\bibfnamefont {Y.}~\bibnamefont
  {Aoki}}, \bibinfo {author} {\bibfnamefont {S.}~\bibnamefont {Bors{\'a}nyi}},
  \bibinfo {author} {\bibfnamefont {S.}~\bibnamefont {Durr}}, \bibinfo {author}
  {\bibfnamefont {Z.}~\bibnamefont {Fodor}}, \bibinfo {author} {\bibfnamefont
  {S.~D.}\ \bibnamefont {Katz}}, \bibinfo {author} {\bibfnamefont
  {S.}~\bibnamefont {Krieg}},\ and\ \bibinfo {author} {\bibfnamefont {K.~K.}\
  \bibnamefont {Szab{\'o}}},\ }\href
  {https://doi.org/10.1088/1126-6708/2009/06/088} {\bibfield  {journal}
  {\bibinfo  {journal} {JHEP}\ }\textbf {\bibinfo {volume} {06}},\ \bibinfo
  {pages} {088}},\ \Eprint {https://arxiv.org/abs/0903.4155} {arXiv:0903.4155
  [hep-lat]} \BibitemShut {NoStop}%
\bibitem [{\citenamefont {Kogut}\ and\ \citenamefont
  {Sinclair}(2002)}]{Kogut:2002zg}%
  \BibitemOpen
  \bibfield  {author} {\bibinfo {author} {\bibfnamefont {J.~B.}\ \bibnamefont
  {Kogut}}\ and\ \bibinfo {author} {\bibfnamefont {D.~K.}\ \bibnamefont
  {Sinclair}},\ }\href {https://doi.org/10.1103/PhysRevD.66.034505} {\bibfield
  {journal} {\bibinfo  {journal} {Phys. Rev. D}\ }\textbf {\bibinfo {volume}
  {66}},\ \bibinfo {pages} {034505} (\bibinfo {year} {2002})},\ \Eprint
  {https://arxiv.org/abs/hep-lat/0202028} {arXiv:hep-lat/0202028} \BibitemShut
  {NoStop}%
\bibitem [{\citenamefont {Giordano}\ \emph
  {et~al.}(2020{\natexlab{c}})\citenamefont {Giordano}, \citenamefont {Kapas},
  \citenamefont {Katz}, \citenamefont {Nogradi},\ and\ \citenamefont
  {Pasztor}}]{Giordano:2019gev}%
  \BibitemOpen
  \bibfield  {author} {\bibinfo {author} {\bibfnamefont {M.}~\bibnamefont
  {Giordano}}, \bibinfo {author} {\bibfnamefont {K.}~\bibnamefont {Kapas}},
  \bibinfo {author} {\bibfnamefont {S.~D.}\ \bibnamefont {Katz}}, \bibinfo
  {author} {\bibfnamefont {D.}~\bibnamefont {Nogradi}},\ and\ \bibinfo {author}
  {\bibfnamefont {A.}~\bibnamefont {Pasztor}},\ }\href
  {https://doi.org/10.1103/PhysRevD.101.074511} {\bibfield  {journal} {\bibinfo
   {journal} {Phys. Rev. D}\ }\textbf {\bibinfo {volume} {101}},\ \bibinfo
  {pages} {074511} (\bibinfo {year} {2020}{\natexlab{c}})},\ \bibinfo {note}
  {[Erratum: Phys.Rev.D 104, 119901 (2021)]},\ \Eprint
  {https://arxiv.org/abs/1911.00043} {arXiv:1911.00043 [hep-lat]} \BibitemShut
  {NoStop}%
\bibitem [{\citenamefont {Mukherjee}\ \emph {et~al.}(2022)\citenamefont
  {Mukherjee}, \citenamefont {Rennecke},\ and\ \citenamefont
  {Skokov}}]{Mukherjee:2021tyg}%
  \BibitemOpen
  \bibfield  {author} {\bibinfo {author} {\bibfnamefont {S.}~\bibnamefont
  {Mukherjee}}, \bibinfo {author} {\bibfnamefont {F.}~\bibnamefont
  {Rennecke}},\ and\ \bibinfo {author} {\bibfnamefont {V.~V.}\ \bibnamefont
  {Skokov}},\ }\href {https://doi.org/10.1103/PhysRevD.105.014026} {\bibfield
  {journal} {\bibinfo  {journal} {Phys. Rev. D}\ }\textbf {\bibinfo {volume}
  {105}},\ \bibinfo {pages} {014026} (\bibinfo {year} {2022})},\ \Eprint
  {https://arxiv.org/abs/2110.02241} {arXiv:2110.02241 [hep-ph]} \BibitemShut
  {NoStop}%
\bibitem [{\citenamefont {Dimopoulos}\ \emph {et~al.}(2022)\citenamefont
  {Dimopoulos}, \citenamefont {Dini}, \citenamefont {Di~Renzo}, \citenamefont
  {Goswami}, \citenamefont {Nicotra}, \citenamefont {Schmidt}, \citenamefont
  {Singh}, \citenamefont {Zambello},\ and\ \citenamefont
  {Ziesch\'e}}]{Dimopoulos:2021vrk}%
  \BibitemOpen
  \bibfield  {author} {\bibinfo {author} {\bibfnamefont {P.}~\bibnamefont
  {Dimopoulos}}, \bibinfo {author} {\bibfnamefont {L.}~\bibnamefont {Dini}},
  \bibinfo {author} {\bibfnamefont {F.}~\bibnamefont {Di~Renzo}}, \bibinfo
  {author} {\bibfnamefont {J.}~\bibnamefont {Goswami}}, \bibinfo {author}
  {\bibfnamefont {G.}~\bibnamefont {Nicotra}}, \bibinfo {author} {\bibfnamefont
  {C.}~\bibnamefont {Schmidt}}, \bibinfo {author} {\bibfnamefont
  {S.}~\bibnamefont {Singh}}, \bibinfo {author} {\bibfnamefont
  {K.}~\bibnamefont {Zambello}},\ and\ \bibinfo {author} {\bibfnamefont
  {F.}~\bibnamefont {Ziesch\'e}},\ }\href
  {https://doi.org/10.1103/PhysRevD.105.034513} {\bibfield  {journal} {\bibinfo
   {journal} {Phys. Rev. D}\ }\textbf {\bibinfo {volume} {105}},\ \bibinfo
  {pages} {034513} (\bibinfo {year} {2022})},\ \Eprint
  {https://arxiv.org/abs/2110.15933} {arXiv:2110.15933 [hep-lat]} \BibitemShut
  {NoStop}%
\bibitem [{\citenamefont {P\'asztor}\ \emph {et~al.}(2021)\citenamefont
  {P\'asztor}, \citenamefont {Sz\'ep},\ and\ \citenamefont
  {Mark\'o}}]{Pasztor:2020dur}%
  \BibitemOpen
  \bibfield  {author} {\bibinfo {author} {\bibfnamefont {A.}~\bibnamefont
  {P\'asztor}}, \bibinfo {author} {\bibfnamefont {Z.}~\bibnamefont {Sz\'ep}},\
  and\ \bibinfo {author} {\bibfnamefont {G.}~\bibnamefont {Mark\'o}},\ }\href
  {https://doi.org/10.1103/PhysRevD.103.034511} {\bibfield  {journal} {\bibinfo
   {journal} {Phys. Rev. D}\ }\textbf {\bibinfo {volume} {103}},\ \bibinfo
  {pages} {034511} (\bibinfo {year} {2021})},\ \Eprint
  {https://arxiv.org/abs/2010.00394} {arXiv:2010.00394 [hep-lat]} \BibitemShut
  {NoStop}%
\bibitem [{\citenamefont {Haque}\ and\ \citenamefont
  {Strickland}(2021)}]{Haque:2020eyj}%
  \BibitemOpen
  \bibfield  {author} {\bibinfo {author} {\bibfnamefont {N.}~\bibnamefont
  {Haque}}\ and\ \bibinfo {author} {\bibfnamefont {M.}~\bibnamefont
  {Strickland}},\ }\href {https://doi.org/10.1103/PhysRevC.103.L031901}
  {\bibfield  {journal} {\bibinfo  {journal} {Phys. Rev. C}\ }\textbf {\bibinfo
  {volume} {103}},\ \bibinfo {pages} {031901} (\bibinfo {year} {2021})},\
  \Eprint {https://arxiv.org/abs/2011.06938} {arXiv:2011.06938 [hep-ph]}
  \BibitemShut {NoStop}%
\bibitem [{\citenamefont {Fukushima}(2008)}]{Fukushima:2008wg}%
  \BibitemOpen
  \bibfield  {author} {\bibinfo {author} {\bibfnamefont {K.}~\bibnamefont
  {Fukushima}},\ }\href {https://doi.org/10.1103/PhysRevD.77.114028} {\bibfield
   {journal} {\bibinfo  {journal} {Phys. Rev. D}\ }\textbf {\bibinfo {volume}
  {77}},\ \bibinfo {pages} {114028} (\bibinfo {year} {2008})},\ \bibinfo {note}
  {[Erratum: Phys.Rev.D 78, 039902 (2008)]},\ \Eprint
  {https://arxiv.org/abs/0803.3318} {arXiv:0803.3318 [hep-ph]} \BibitemShut
  {NoStop}%
\bibitem [{\citenamefont {Kov\'acs}\ \emph {et~al.}(2016)\citenamefont
  {Kov\'acs}, \citenamefont {Sz\'ep},\ and\ \citenamefont
  {Wolf}}]{Kovacs:2016juc}%
  \BibitemOpen
  \bibfield  {author} {\bibinfo {author} {\bibfnamefont {P.}~\bibnamefont
  {Kov\'acs}}, \bibinfo {author} {\bibfnamefont {Z.}~\bibnamefont {Sz\'ep}},\
  and\ \bibinfo {author} {\bibfnamefont {G.}~\bibnamefont {Wolf}},\ }\href
  {https://doi.org/10.1103/PhysRevD.93.114014} {\bibfield  {journal} {\bibinfo
  {journal} {Phys. Rev. D}\ }\textbf {\bibinfo {volume} {93}},\ \bibinfo
  {pages} {114014} (\bibinfo {year} {2016})},\ \Eprint
  {https://arxiv.org/abs/1601.05291} {arXiv:1601.05291 [hep-ph]} \BibitemShut
  {NoStop}%
\bibitem [{\citenamefont {Isserstedt}\ \emph {et~al.}(2019)\citenamefont
  {Isserstedt}, \citenamefont {Buballa}, \citenamefont {Fischer},\ and\
  \citenamefont {Gunkel}}]{Isserstedt:2019pgx}%
  \BibitemOpen
  \bibfield  {author} {\bibinfo {author} {\bibfnamefont {P.}~\bibnamefont
  {Isserstedt}}, \bibinfo {author} {\bibfnamefont {M.}~\bibnamefont {Buballa}},
  \bibinfo {author} {\bibfnamefont {C.~S.}\ \bibnamefont {Fischer}},\ and\
  \bibinfo {author} {\bibfnamefont {P.~J.}\ \bibnamefont {Gunkel}},\ }\href
  {https://doi.org/10.1103/PhysRevD.100.074011} {\bibfield  {journal} {\bibinfo
   {journal} {Phys. Rev. D}\ }\textbf {\bibinfo {volume} {100}},\ \bibinfo
  {pages} {074011} (\bibinfo {year} {2019})},\ \Eprint
  {https://arxiv.org/abs/1906.11644} {arXiv:1906.11644 [hep-ph]} \BibitemShut
  {NoStop}%
\bibitem [{\citenamefont {Gao}\ and\ \citenamefont
  {Pawlowski}(2021)}]{Gao:2020fbl}%
  \BibitemOpen
  \bibfield  {author} {\bibinfo {author} {\bibfnamefont {F.}~\bibnamefont
  {Gao}}\ and\ \bibinfo {author} {\bibfnamefont {J.~M.}\ \bibnamefont
  {Pawlowski}},\ }\href {https://doi.org/10.1016/j.physletb.2021.136584}
  {\bibfield  {journal} {\bibinfo  {journal} {Phys. Lett. B}\ }\textbf
  {\bibinfo {volume} {820}},\ \bibinfo {pages} {136584} (\bibinfo {year}
  {2021})},\ \Eprint {https://arxiv.org/abs/2010.13705} {arXiv:2010.13705
  [hep-ph]} \BibitemShut {NoStop}%
\bibitem [{\citenamefont {Bernhardt}\ \emph {et~al.}(2021)\citenamefont
  {Bernhardt}, \citenamefont {Fischer}, \citenamefont {Isserstedt},\ and\
  \citenamefont {Schaefer}}]{Bernhardt:2021iql}%
  \BibitemOpen
  \bibfield  {author} {\bibinfo {author} {\bibfnamefont {J.}~\bibnamefont
  {Bernhardt}}, \bibinfo {author} {\bibfnamefont {C.~S.}\ \bibnamefont
  {Fischer}}, \bibinfo {author} {\bibfnamefont {P.}~\bibnamefont
  {Isserstedt}},\ and\ \bibinfo {author} {\bibfnamefont {B.-J.}\ \bibnamefont
  {Schaefer}},\ }\href {https://doi.org/10.1103/PhysRevD.104.074035} {\bibfield
   {journal} {\bibinfo  {journal} {Phys. Rev. D}\ }\textbf {\bibinfo {volume}
  {104}},\ \bibinfo {pages} {074035} (\bibinfo {year} {2021})},\ \Eprint
  {https://arxiv.org/abs/2107.05504} {arXiv:2107.05504 [hep-ph]} \BibitemShut
  {NoStop}%
\end{thebibliography}
\providecommand{\noopsort}[1]{}\providecommand{\singleletter}[1]{#1}%

\end{document}